\documentclass[review]{elsarticle}

\usepackage{ntheorem}
\usepackage{amsmath}{
      \newtheorem*{theorem-no}{$h_0$:}
  }

\theoremstyle{definition}

\usepackage[margin=2cm]{geometry}
\usepackage{soul}
\usepackage{amssymb}
\usepackage{amsmath}
\usepackage{wrapfig} 
\usepackage{multirow}
\usepackage{colortbl}
\usepackage{adjustbox}
\usepackage{longtable}
\usepackage{graphicx}
\usepackage{booktabs}
\usepackage{rotating}
\usepackage{array}
\newcolumntype{C}[1]{>{\centering\arraybackslash}m{#1}}
\newcolumntype{N}[1]{>{\raggedright\arraybackslash}m{#1}}
\usepackage{lscape} 

\usepackage{multirow}
\usepackage{subfig}

\usepackage{hyperref}

\journal{ArXiv}

\begin{document}

\begin{frontmatter}

\title{Understanding and Assessment of Mission-Centric Key Cyber Terrains for joint Military Operations}

\author[mysecondaryaddress]{\'{A}lvaro Luis Mart\'{i}nez\corref{mycorrespondingauthor}}
\ead{aluism@indra.es}

\author[mysecondaryaddress]{Jorge {Maestre Vidal}\corref{mycorrespondingauthor}}
\cortext[mycorrespondingauthor]{Corresponding author}
\ead{jmaestre@indra.es}

\author[mymainaddress]{V\'{i}ctor A. Villagr\'{a} Gonz\'{a}lez\corref{mycorrespondingauthor}}
\ead{victor.villagra@upm.es}

\address[mysecondaryaddress]{Indra, Digital Labs, Avenida de Bruselas, 35, Alcobendas, 28108 Madrid, Spain}

\address[mymainaddress]{Universidad Polit\'ecnica de Madrid. ETSI de Telecomunicaci\'on. Departamento de Ingenier\'ia de Sistemas Telem\'aticos, Madrid, Spain}

\begin{abstract}
Since the cyberspace consolidated as fifth warfare dimension, the different actors of the defense sector began an arms race toward achieving cyber superiority, on which research, academic and industrial stakeholders contribute from a dual vision, mostly linked to a large and heterogeneous heritage of developments and adoption of civilian cybersecurity capabilities. In this context, augmenting the conscious of the context and warfare environment, risks and impacts of cyber threats on kinetic actuations became a critical rule-changer that military decision-makers are considering. A major challenge on acquiring mission-centric Cyber Situational Awareness (CSA) is the dynamic inference and assessment of the vertical propagations from situations that occurred at the mission supportive Information and Communications Technologies (ICT), up to their relevance at military tactical, operational and strategical views.  In order to contribute on acquiring CSA, this paper addresses a major gap in the cyber defence state-of-the-art: the dynamic identification of Key Cyber Terrains (KCT) on a mission-centric context. Accordingly, the proposed KCT identification approach explores the dependency degrees among tasks and assets defined by commanders as part of the assessment criteria. These are correlated with the discoveries on the operational network and the asset vulnerabilities identified thorough the supported mission development. The proposal is presented as a reference model that reveals key aspects for mission-centric KCT analysis and supports its enforcement and further enforcement by including an illustrative application case.
\end{abstract}


\end{frontmatter}


\section{Introduction}
The cyberspace is defined as the global domain consisting of all interconnected communication, information technology and other electronic systems, networks and their data; recently being consolidated as fifth domain in the modern theater of warfare, joining land, sea, air and space  \cite{refMe-Sal4}. There, Joint Functions (JFs) like cyber maneuvers, fires, Command and Control (C2), intelligence, information, sustainment, or force protection are enforced in the context of Defensive Cyberspace Operations (DCOs) and /or Offensive Cyberspace Operations (OCOs); supporting or supported by kinetic domain actuations. Despite these dependencies, the cyberspace is completely different from kinetic domain mainly because a man-made nature, being partially nonphysical (digital) and exempt of conventional geographical boundaries \cite{newTDoS}. It is described by the high reachability of CIS (Communication and Information System) assets, the short time effects on cyber Courses of Action (CoAs) against their large preparatory time, the growing asymmetrically of their impact (typically vertically/propagating to collateral hybrid dimensions), or their intangibility; the later making difficult the cyber damage assessment calculation \cite{newCSA1}.

In contrast with state-of-the-art dual use cybersecurity enablers, raw cyber defence effectors suited for military operations shall embrace a mission-centric vision, where cyber assessment and decisions must be complaint with the military operational context, including among other mission goals and tasks on which the cyber actuations act, their interdependencies, phasing, joint/combined actuations, vertical propagations between technical, tactical, operational and strategic levels, etc. This demands a clear transposition between raw impact on Communication and Information System (CIS) dimensions (typically Confidentiality, Integrity and Availability) up to their mission-level implications, being examples of the latter the potentially consequent delays in the execution of tasks, loss of surprise factor, or reduction of the commander's agility in making new decisions and planning CoAs \cite{ares1}. 

Accordingly, an essential aspect on the effectiveness of the capabilities for supporting cyber defence operations is their ability to consider which mission dependencies are essential in order to adequately valuate/assess each cybernetic or procedural asset, and the dynamically change of the valuation as the mission progress. In this context, Key Cyber Terrains (KCTs) are defined as \textit{systems, devices, protocols, data, software, processes, cyber personas, and other networked entities that comprise, supervise, and control cyberspace} \cite{ares1}, posing military advantage, and if jeopardized, potentially causing mission failure \cite{ares2}. But despite the relevance of the KCT concept, it was vaguely publicly explored by the research community, where cultural and cross-domain misunderstanding has led to ambiguities and misperception, typically approached from an excessively civilian perspective; and mostly overlooking its mission-centric implications. 

With the motivation of contributing to the research on mission-centric KCT discovery and assessment, this papers reviews and in-depth analyzes the KCT concept and its implications. This is conducted thorough a Cyber Situational Awareness (CSA) perspective, and under the assumption that its relevance will increasingly grows towards achieving and accurate cross-domain Common Operational Picture (COP) \cite{refMe-Sal1}. The conducted research extends the work preliminary presented to the research community and cyber defence practitioners in \cite{newAresKCT}, compiling the widely received feedback and increasing its original scope from raw technological aspects up to the KCT mission implications. Given the high interest it aroused, the paper strengthens the introduced KCT conceptualization, extends the KCT discovery and assessment reference model, details its mission-centric implications and provides an extended description of the analytical and empirical evaluations. The following enumerates the main contributions of the conducted research.

\begin{itemize}
\item The paper presents an in-depth review of the current landscape on cyber defense and mission-centric cyber risk management, emphasizing the existing efforts towards KCT analysis.
\item A reference framework for dynamically recognize critical assets on the cyberspace is proposed, which reveals a subset of key pillars that may guide further research actions.
\item The traceability between widely adopted kinetic terrain factors and their transposition to military operations on the cyberspace, is explored.
\item The proposal reviews the DOTMLPF-I (Doctrine, Organization, Training, Materiel, Logistics, Personnel, Facilities, and Interoperability) dimensions for KCT assessment related capability development.
\item The application of the proposal for KCT analysis at offensive and defensive military thinking is discussed
\item The proposal has been instantiated and analytically validated under an illustrative use case, which details all the required data processing activities.

\end{itemize}
This paper is organized into seven sections, being the first of them the present Introduction. Section II reviews the state-of-the art on mission-centric cyber defence and KCT assessment. Section III presents the conducted research design principles. Section IV introduces a novel dynamic KCT identification scheme. Section V analyzes the foreseen KCT assessment capability development dimensions. This section also discussed the proposal application at both offensive and defensive thinking. Section VI details the proposal application on a case of study. Finally, Section VII presents the achieved conclusions and suggestions for future work,

\section{Background}
\label{Sec2}
In recent decades a plethora of Risk Management (RM) and impact analysis frameworks have risen up in order to prompt the organizational awareness against cyber threats on which their business depends, leading from a better understanding that range from revealing potential attack paths and vulnerability mitigation plans, up to the analysis of their operational consequences. In order to facilitate the understanding of the conducted research, the following briefly reviews the state-of-the-art on cyber risk management, their relationship with military solutions for acquiring mission-centric Cyber Situational Awareness (CSA), and the existing capabilities for cyber asset assessment; the latter including the discrimination of the essential cyber terrains for military actions.

\subsection{Cyber Risk Management}
According to the National Institute of Standards and Technology (NIST), a Risk Management Framework (RMF)  is a process to integrate security and risk management capabilities \cite{ares4}. They comprise six major stages: 1) preparing the organization towards managing risks; 2) categorizing the adopted information systems; 3) selecting the most suitable security controls; 4) security controls implementation and documentation; 5) security control assessment; authorization the information system operation; and 6) continuous monitoring the security controls effectiveness, as well as any system-level or environmental variation \cite{refMe-new6}. Beyond the NIST RMF approach, the cybersecurity community has joint efforts towards developing, integrating and validating complementary risk management solutions, which are tailored to the different operational and environmental contexts. This result in alternative widely accepted and implemented frameworks, as is the case, among others, of NIST 800-30, OCTAVE Allegro, ISO 31000, ISO 27005, MEHARI, or Magerit. 

With a heavy focus on guiding the development of Risk Assessment (RA) actions, the NIST 800-30 reference model highlights four mandatory steps \cite{ares5}: 1) to scope the risk assessment process by identifying the model application context \cite{refNIst-stage1}; 2) in order to calculate security risks assuming their uncertainly degree, the potential threats and vulnerabilities shall be identified, including their related impact and occurrence likelihood \cite{refNIst-stage2}; 3) in order to facilitate decision-making and countermeasure enforcement, the risk assessment results shall be communicated to the convenient response actors \cite{refNIst-stage3} ; and 4)   to support the related risk management activities by dynamically  monitoring and maintaining the risk assessment procedures \cite{refNIst-stage4}.

Complementary, OCTAVE Allegro (Operationally Critical Threat, Asset, and Vulnerability Evaluation) entails another example of widely adopted risk management framework, which aims on  streamlining and optimizing the process of assessing information security risks so that an organization can obtain sufficient results with a small investment in time, people, and other limited resources \cite{ares6}. This is conducted thorough a preliminary definition of the risk measurement criteria, developing a tailored information asset profile, identifying the organizational information asset containers and their areas of concern, discovering the potential threat scenarios; and from them conduct the risk analysis and selection of most suitable mitigation actions.

The International Organization for Standardization (ISO) published its principles and generic guidelines on managing risks faced by organizations through its 31000 family of standards \cite{ares7}; which defined risk management as an integral process to be enforced by the protected organizations. In the ISO 27005 \cite{ares8} the risk management standards and guidelines are adapted to the Information System Security particularities. As answer to ISO/IEC 27005:2011 guidelines, proposals like
MEHARI (Method for Harmonized Analysis of Risk) \cite{ares9} facilitate an organization compliance for their Information Security Risk Management (ISRM) processs (e.g., ISO 27001 2013 revision), providing both documental supports and practical tools for their enforcement.

Another world-wide spread AR and RM methodology, with significant relevance for the military sector (e.g., NATO) is MAGERIT \cite{ares10}. Developed by the Spanish Public Administration and supported by the EAR/PILAR linked software, MAGERIT guides the ISO 31000 adoption and implementation based on each organization particularities. MAGERIT is based on six steps: to characterize assets and their inter-dependencies; to identify threats; to estimate the impact; to estimate the risk defined as the impact weighted by the likelihood of the threat; to select safeguards and measure its effectiveness; and to calculate residual impact and residual risk.

According to \cite{refNIst-stage2}, the risk management approaches can be categorized as qualitative and quantitative, which depends on the nature of the information on which security decisions rely. In this context, the qualitative models focus on the analysis of the likelihood of a specific risk event occurs, as well as its impact on the overall business operations and goals \cite{refQualitativeRM}, being the approaches described above examples of qualitative-base RA/RM solutions. In contrast, and with a much lower presence at design, adoption and standardization/harmonization RA/RM actions, quantitative methods operate on numerical terms where Bayesian Decision Networks (BDN) \cite{refQuantitativeRM-1} guide cybersecurity benefit-cost analytics. Anther examples of quantiative RA/RM are AVARCIBER \cite{refQuantitativeRM-2} for extending the specific parameters of ISO 27005 \cite{ares8}. 

Despite the maturity of these methodologies on civilian applications, none of them has been conceived for supporting mission-centric RM/RA by design, which evidences a huge gap between the military needs and the background delivered by civilian industry and research institutions. Solutions like MAGERIT partially cover needs related with the management of the CIS technological plane, but they do not properly support the assessment of vertical propagation from cyber situations to the mission businesses and goals.

\subsection{Mission-centric Cyber Situational Awareness}
Situational Awareness (SA) is a human ``brain state” that refers to being conscious of the operational context and the development of planned/ongoing actions with the aim on selecting and planning more effective reactive/proactive CoAs. This term has been actively revisited by Mica Endsley, who proposed the most adopted SA model in recent decades \cite{ares11}. The Endlsey's model layered the SA into three mayor phases: Perception of the operational environment; Comprehension of the perceived information so the inference of new related knowledge is possible; and the Projection of the SA at different future time horizons. Other authors like Bedny and Meister \cite{ares14} and Smith and Hancock \cite{ares15} proposed alternative models of SA, which have been widely modified since their publication. A well-known abstraction of the SA paradigm is the OODA (Observe Orient Decide Act) loop model proposed by former USAF colonel John Boyd \cite{ares16} for supporting fast decision-making, actually constituting the core pillar of the C2 solutions.

By following the coining SA philosophy, the research community has put significantly effort towards adapting these cognitive models to the cyberspace, referring to the resultant mind state as Cyber Situation Awareness (CSA). As Jajodia et al. stated in \cite{ares17}: \textit{to protect critical network infrastructures and missions, we must understand not only the vulnerabilities of each individual system, but also their inter-dependencies and how they support missions}, which gains difficulty when operating on emerging technological ecosystems \cite{refMe-new7,refMe-Sal3}, combining the perception of both insider and outsider threats \cite{refMe-new3} or facing adversarial evasion tactics \cite{refMe-new1, refMe-Sal2} . At the same time, they proposed a framework to obtain mission-centric SA (Cauldron) combining data fusion, network paths of vulnerabilities, alert correlation, mission impact analysis and recommended reactive/proactive mitigation actions \cite{refMe-new5,refMe-new8}.  Based on the three Endsley's SA levels, McGuinness and Foy \cite{ares18} added a fourth layer towards developing an alternative CSA model grounded on Perception, Comprehension, Projection and Resolution.  On the other hand, Lenders et al. \cite{ares19} made evolve the John Boyd's OODA loop to a cyber perspective. Buckshaw et al. \cite{ares20} proposed MORDA (Mission Oriented Risk and Design Analysis of Critical Information) based on attack trees, adversary models, user models, service provider models and analysis models to define a quantitative risk assessment and management by means of Multiple Objective Decision Analysis algorithms and SMEs (Subject Matter Expert).

As response to the increasing need for standardized CSA solutions, MITRE developed a related framework fitted to the NATO Communications and Information Agency \cite{ares21} which comprised Threat Intelligence, Dependency and Impact Analysis, Analysis of Alternatives and Emerging Solutions module to be aware of threats and actors, dependencies and possible countermeasures against discovered threats. Public convergence actions, as is the case of the project H2020 PROTECTIVE, are exploring the context awareness to assess assets critically, the later relying on three basic features: Mission Impact Management, Asset State Management, and Mission and Asset Information Repositories. Complementary, research like \cite{ares22} is exploring the definition of intelligence systems for critical infrastructure protection under hybrid scenarios (physical and cyber data sources), the latter by taking advantage of three main modules: Data Gathering, Data Analysis and Data Visualization. Finally, in \cite{ares23} Franke and Brynielsson analyzed both CSA approaches: on one hand industry control and critical infrastructure and in other hand military. They concluded that first approach has been widely researched whereas military CSA application has much less bibliographical coverage.

\subsection{Key Cyber Terrains}
Occidental military doctrines define a Key Terrain as \textit{any locality, or area, the seizure or retention of which affords a marked advantage to either combatant} \cite{ares24}. This definition aligns with the SA purpose of being aware of the environment and hostile actors in order to take some advantage to accomplish each mission goal. When developing CSA, the Key Terrain concept embraces the cyber domain with the aim on identifying those cyber-dependant components that play crucial roles for a mission development advantage. They are refered to as Key Cyber Terrains (KCT). But in contrast with conventional kinetic terrains, KCTs are more difficult not only to defend but also to identify due to their ICT (Information and Communication Technology) nature \cite{ares1}, which affects reachability. For example, the location and temporal components that characterize physical environments might be completely vanished in the digital World, where hostile actors may remotely gain control of critical infrastructure from thousand kilometers, change systems policies in milliseconds and use advanced concealing techniques to avoid being discovered by conventional Intelligence, Surveillance and Reconnaissance (ISR) procedures.

Consequently, the cyberspace entails a fast and dynamic environment with a fuzzy sense of control. Therefore its analysis should be addressed with a more complex and dynamic approach. For example, in \cite{ares25} is suggested a KCT identification framework based on the traditional Key Terrain analysis OCOKA (Observation and Fields of Fire, Cover and Concealment, Obstacles, Key Terrain and Avenues of approach) \cite{ares27}. This KCT framework deals with the identification of the potential exploitable assets; identification of avenues of approach, that is, attack vectors; and the deployment of obstacles and camouflage techniques to limit avenues of approach. In \cite{ares28} five levels where KCT can be identified were defined, thus extending those preliminary declared by the US DoD (Department of Defense) on kinetic encounters: supervisor, cyberperson, logical, physic and geographical.

Price et al. \cite{ares29} suggested a mission-centered criticallity and dependency analysis to assist defence analysts to acquire CSA via Multi-Attribute Decision Making (MADM) guided by the TOPSIS (Technique for Order of Preference by Similarity to Ideal Solution) method. They conclude that families (sets of assets) are more relevant for mission severity than the single asset assessment. In \cite{ares30} three strategies for supporting Cyber-Physical Key Terrains under simulated movement models were proposed: centrality analysis, dynamical systems (Markov Models) and agent-based modelling. They concluded that although centrality and dynamical systems strategies can approach agent-base modelling, the latter shows much more flexibility to represent each use case model. Complementary, the Massachusetts Institute of Technology (MIT) \cite{ares31} remarked the importance of KCT discovery on military missions, and suggested three different for its proper development: manual identification by SMEs; logs and network capacities analysis; and perturbational analysis.

Beyond the theoretical frameworks and basic empirical demonstrations presented in the bibliography, the research on KCT discovery and management presents an essential, but rarely explored topic. Few proposals addressed the KCT related challenges from a mission-centric perspective, most of them heavily relying on up-to-down procedures that difficulty adapt to the dynamism required on real operational environments. On the other hand, they usually are not linked to mission assessment capabilities, so the vertically propagated impact on a KCT (technical war plane) to the upper war levels (tactical, operational, strategy) have been barely taken into consideration; in this way, failing when correlating the impact on KCTs regarding the mission goals.

\section{Design Principles}
\label{Sec3}
This section describes the main consideration taken into account during the conducted research, highlighting objectives, assumptions and limitations.

\subsection{Objectives}
The principal objective of the conducted research has been to design a system able to dynamically discover and score assets deployed on the cyberspace considering the propagated criticality of them regarding the mission tasks. As secondary goals, it was stated that: 1) when calculating the criticality score, the proposal shall be able to consider the potential horizontal/vertical propagation between task and assets dependencies; 2) the assessment process shall be continuous and dynamic, since it is assumed that the ICT ecosystem on which the mission depends evolves over time; 3) the outcomes of the proposal shall be clear and understandable, so they can contribute to the acquisition of cyber situational awareness.

\subsection{Assumptions and requirements}
The design requirements to develop the research are stated as follows:
\begin{itemize}
\item The proposal implementation will require to be able to reach the analyzed assets, since it shall relay on well-known network discovery and port scanning applications.
\item In order to perform a deep active scanning against cyber assets, the KCT discovery system requires to have enough privileges. This is due that it must be allowed to perform advanced scanning options.
\item The network scanning results are stored into the system by the scanning procedures as XML files (or other common data representation formats). No data treatment policies were applied on them (e.g., anonymization, filtering, aggregation, etc.), so they can be processes in raw for further analytic actions. 
\end{itemize}

\subsection{Limitations}
Some limitations have been foreseen at the early stage of the research, or during the core research actions, highlighting:
\begin{itemize}
\item The Concept of Operation (COP) as a set of missions is not supported in the current system development state. In further more mature versions it shall be properly aligned with the military standards and the Joint Operations doctrines.
\item The proposed model does not implement threats, risks nor impacts assessment process. In further developments, the system would work with these models due to the integration with impact assessment and risk management modules.
\item The proposal configuration heavily relies on a top-to-down mission planning (similar to those in the state of the art, and as suggested by the doctrines). It is expected that alternative approaches may imply a higher degree of automatism (e.g., machine learning, prediction, etc.), which preliminary were out of the scope of the conducted research due to the doctrinal barriers inherent in adopting unsupervised or semi-supervised capabilities on military enablers.
\end{itemize}

\section{Dynamic Key Cyber Terrain identification scheme}
The following describes a novel methodology for mission-centric KCT discovery, emphasizing its core elements, the KCT discovery procedure steps, and the scoring system applied for inferring the criticality of each analyzed cyber asset.

\subsection{KCT Discovery Components}
\begin{figure*}[t!]
  \centering
  \includegraphics[width=0.6\linewidth]{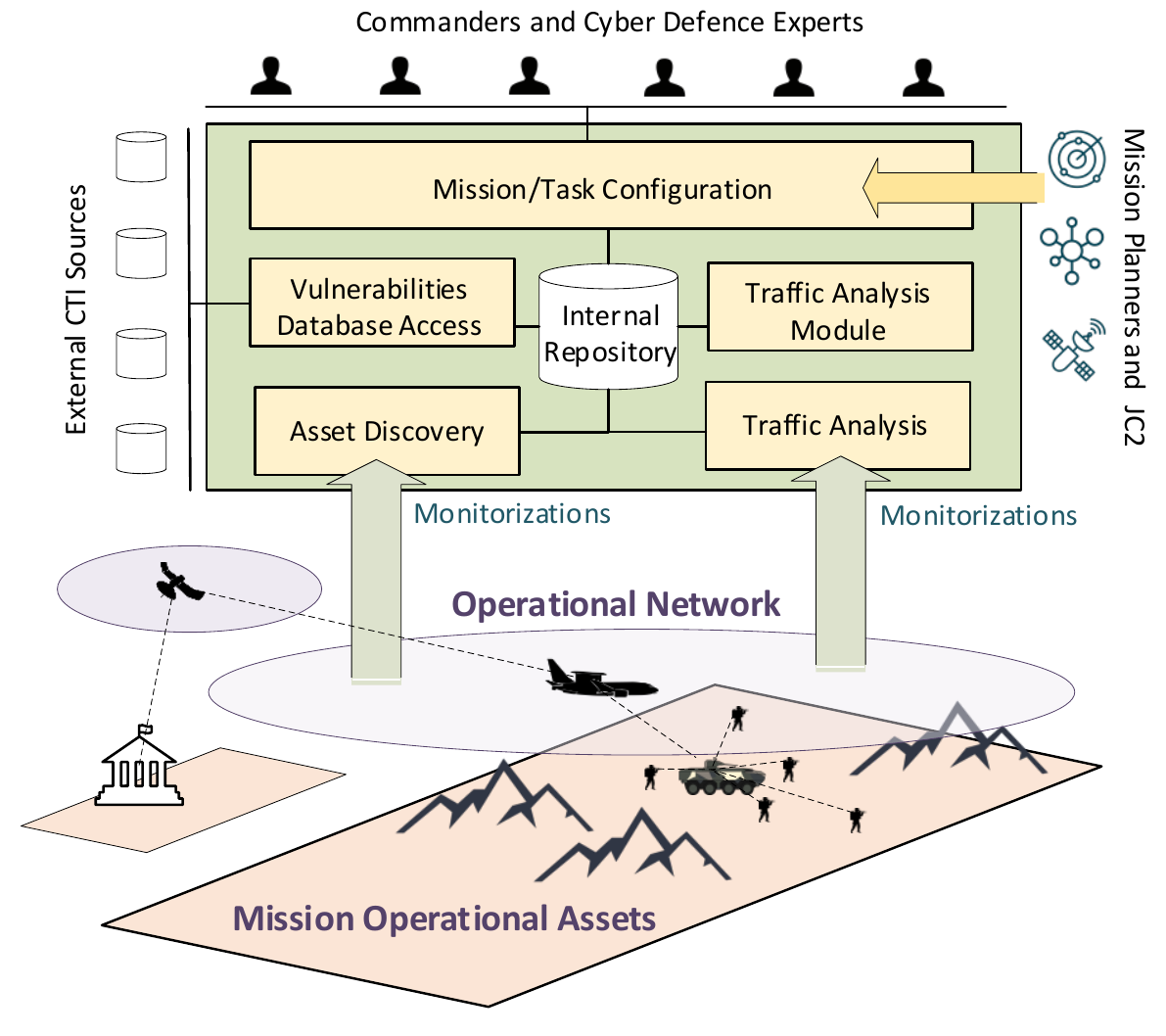}
  \caption{\label{fig1} Architecture overview}
\end{figure*}

In the grounds of the design principles described in Section 3, the following introduces a novel KCT identification system able to dynamically value each cyber terrain. This is dynamically conducted from a mission-centric perspective, by design assuming its application for supporting the acquisition of cyber situational awareness.
The proposal is decoupled into the following core modules as represented in Fig. \ref{fig1}:

\subsubsection{Mission/Task Configuration Module}
The Mission/Task Configuration Module (MTCM) feeds the KCT assessment functionalities with information provided by mission planning enablers, Joint Command and Control (JC2) systems and/or any other capability able to enrich the mission-centric vision of the operational environment \cite{refMissionPlanner}. Beyond linking theses capabilities with the rest of the knowledge management effectors, the MTCM is able to operate on an own abstraction of the mission topology, thus reducing the dependencies with external functionalities to only communication of significant events concerning changes on the mission map (e.g., status of the mission lines of development, re-tasks, enforced CoAs, etc.). This abstraction relies on dynamic sequences of tasks and phases represented via Business Process Model and Notation (BPMN) \cite{refBPMN}, so it can be easily transposed to standardized representations, as is the case of the the Military Scenario Definition Language (MSDL), Simulation Interoperability Standards Organization (SISO) \cite{refSISO} or dual-use approximations to modeling businesses for growing technological enablers \cite{ref5gmodel}. Note that these abstractions are not mandatory for the proper functioning of the proposed KCT Assessment method, but they significantly enhance its interoperability. On these grounds, the MTCM models the tasks, CIS-related operational assets, their relationships, dependency degrees and planned/ongoing tasks' severity. From this, the KCT Discovery Module (KDM) will compute the mission-level severity propagation of the impact of cyber threats, between operational dependencies 

\subsubsection{Assets Discovery Module}
The Assets Discovery Module (ADM) comprises the functionalities related with automatically identifying and mapping the cyber assets spotted on the mission operational network, in order to have a complete knowledge about physical and virtual infrastructure on which the success on the planned/ongoing mission relies on. On these grounds, the ADM main purpose is to automatically discover cyber assets in the mission operational network, like devices or services running on the network through active port scanning. In order to be successfully identified, a CPE (Common Platform Enumeration) should be extracted, which allows to use Vulnerabilities Database Access Module (VDAM) for retrieving CVSS asset scores. Discovery events are reported by cyber sensors and telemetry services grouped into four classes \cite{refNIst-stage4}: discovery, removal, modification and notification. 

The Discovery events report situations related to incorporation of new assets, as is the case of instantiating of new Virtual Network Functions (VNFs) or deploying new nodes. The ADM updates its asset inventory each time a Discovery event is communicated. As opposite, Removal events report situations related with their deletion, which may be triggered by multiple causes ranging from intentional elimination of virtual resources up to the lost of assets due to hostile actions. Each asset deletion implies its removal from the asset inventory. Since conducting misison-centric KCT assessments requires that the ADM registers each asset modification (not only instantiation or removal) that may impact on its value for the mission purpose, the Modification events continuously report and refresh the ADM inventory based on changes like variations of the asset location, software updated, etc. Finally, the Notification events report to the ADM any other discoverable situation related with the assets that may be useful for their valuation, as is the case of the presence of unused resources or detection of requests for special configurations.

\subsubsection{Traffic Analysis Module} 
The Traffic Analysis Module (TAM) is in charge of monitoring and analyzing traffic from the operational network in order to infer CIS-level characteristic of the mission development that are not noticeable by the discovery reports managed by the ADM. The TAM monitoring capabilities may deploy active or passive measurement tools able to capture the traffic crossing monitoring points (e.g., packet-based, flow-based, etc.) and/or host-based linked information; and export them to standardized models that facilitate further management and analysis \cite{refMonitoring}. These capabilities may be complemented by cyber sensors like Network-based Intrusion Detection Systems (NIDS), Host-based Intrusion Detection Systems (HIDS) or SIEM (Security Information and Event Management). In complex operational contexts, the perception of the CIS environment may be supported by external Network Operations Center (NOCs) or Security Operation Center (SOCs) \cite{refNoC}. It is worth to highlight that as a first implementation of the proposed solution, the conducted experimentation implemented basic telemetry services for network monitoring, which exported packet-based observations as Pcap files. They allowed to assess whether or not each communication between assets is encrypted by using protocols like TLS (Transport Layer Security), as wells as to determine if an asset requires high availability. The Analytics conducted by the TAM may vary according to the operational context and the analytical resource availability, typically addressing traffic prediction, traffic classification, failure management or detection/correlation of cyber threats. It may aim on reactive or proactive targets, the second facilitating the identification and assessment of vertical/horizontal propagation of situations topologically, but also in time (thus supporting the decision of anticipative CoAs) \cite{refMe-new5}.

\subsubsection{Vulnerabilities Database Access Module} 
As highligthed in \cite{refNew-CTIGathering} the new generation of cyber threats are rapidly evolving, resembling multi-vectored and often multi-staged \textit{modus operandi}. The first refers to the possibility of horizontal/vertical propagation towards infering diverse impacts, ranging from raw CIS-level consequences, up to complex strategical/operational hybrid consequences across the Political, Military, Economic, Social, Informational and Infrastructure (PMESII) spectrum \cite{new-hybridDamage}. On the other hand, multi-staged threats can be understood from the defensive perspective of a “kill chain” as a sequence of stages required for an attacker to success. The proper detection, assessment and management of these treats require a continuously adapted related knowledge-base, which shall guide the cyber defence automatism and human decision-making. In this context, the Vulnerabilities Database Access Module (VDAM) of the proposal, addressed the activities involved in feeding the rest of the KCT assessment functionalities with the needed Cyber Threat Intelligence (CTI). This component intermediates between external CTI repositories (CERTs, Bug Hunting Platforms, SOCs, etc.), by importing, updating and if needed removing the acquired knowledge. In the context of the conducted experimentation, this module provided an interface to request to NIST's NVD web service the CVEs (Common Vulnerabilities and Exposures) and the CVSS (Common Vulnerability Scoring System) associated to the assets discovered in the operational network.

\subsubsection{KCT Discovery Module (KDM)} 
Based on the feeds of the aforementioned components, and by integrating the mission-level perspective at the analysis of the CIS-level context, the KCT Discovery Module (KDM) conducts the calculations toward assessing the critically of each mission operational asset. This process requires to retrieve topological information as well as traffic and vulnerabilities metrics to obtain a final score for the asset's criticality for both tasks and mission perspective. The result is represented in a dashboard via Graphical User Interface (GUI) for the commander to be dynamically aware of the mission and tasks KCTs in order to make decisions. Since the KDM entails the core capability of the proposed scheme, its will be discussed in more detail at Section 4.2.

\subsubsection{Internal Repository (IR)} 
The Internal Repository (IR) of the proposal implements the capabilities for organizing, storing and retrieval the acquired local knowledge, including its management, verification and update. Given the dynamism inherent in real operational environments, the databases that may comprise the IR shall consider the fast management and retrieval of large information volumes, with high heterogeneity and high dimensionality; if required, distinguishing natural/raw data lakes from persistence of analytical information storage and managers. For simplicity, in the context of the conducted experimentation a relational database was implemented as persistence layer, which stored and facilitates the retrieval of all information needed to compute the final assets score: mission topology, dependencies, severity, vulnerabilities, traffic information, etc. However, and beyond experimental purposes, advanced data management architectures and systems should be considered \cite{ref-PlataformaDato} at real deployment, which shall adopt the data security and classification mechanism that guarantee the confidentiality, integrity, and availability of the information to be analyzed (e.g., NATO STANAG 4774/4778 for security data labeling). In this regard, the defence community is increasingly interested on adopting non perimeter-based data securization (e.g., zero trust, data-centric security, etc.) \cite{ref-newDCS}.

\begin{figure*}[t!]
  \centering
  \includegraphics[width=1\linewidth]{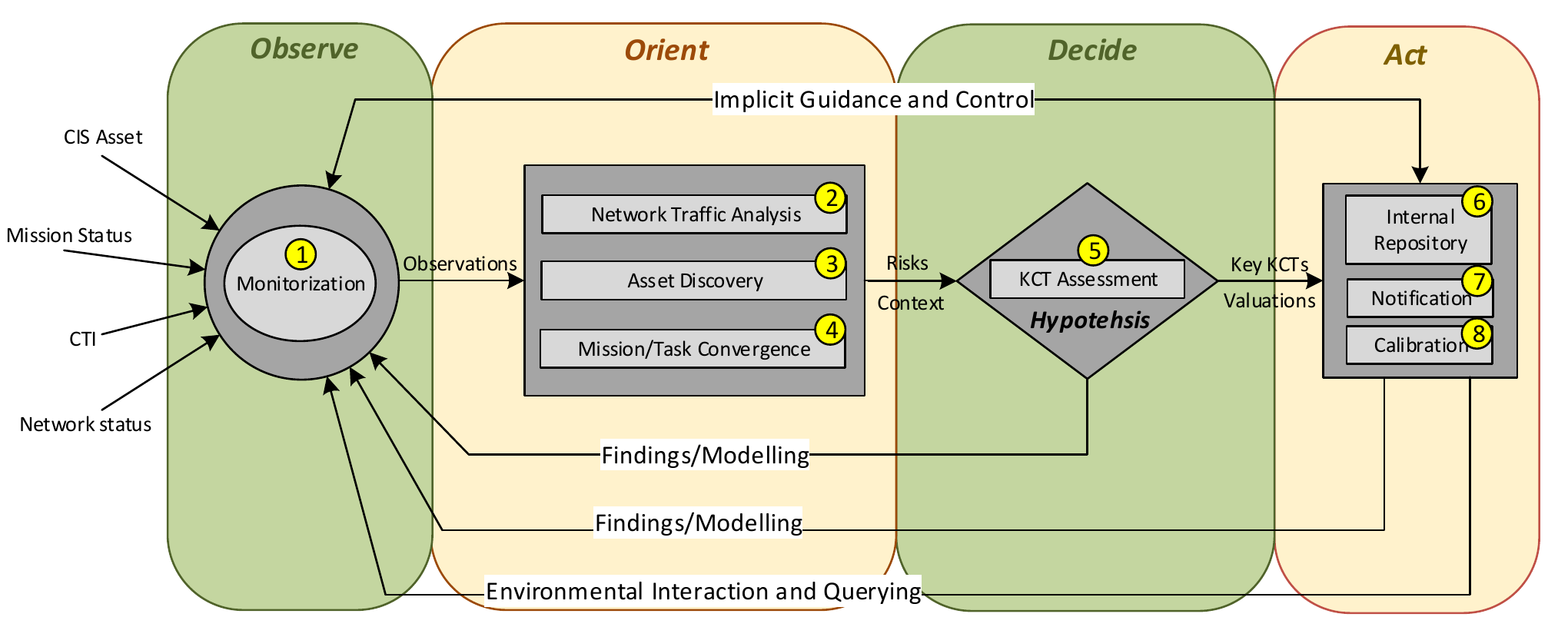}
  \caption{\label{fig2} System flow chart.}
\end{figure*}

\subsection{KCT Discovery Methodology}
The process of discovering KCT entails seven stages that cyclically iterate in convergence with the Observe-Orient-Decide-Act (OODA) loop (see Fig. \ref{fig2}). Introduced by the USAF Colonel John Boyd, the OODA loop was designed aiming on facilitating human decision-making at complex operational circumstances, as is the case of air combat operations \cite{ref-newOODA}. Accordingly, an \textit{Observe} stage analogous to the Perception layer of the Endsley's model \cite{ares11} gathers information from the operational environment, which is where the proposed KCT assessment methodology is feed by CIS-level, mission-level and CTI data sources. Then the \textit{Orient} conducts reasoning actions towards inferring new knowledge from the preliminary one. This is when the main functionalities on ADM, MTCM, TAM operate, creating a baseline that makes converge the acquired CIS-level operational picture regarding the mission-centric context. At the \textit{Decide} stage the KDM functionalities will calculate the mission-centric valuation per CIS asset, revealing the critical terrains that can be tagged as KCT. Finally, at \textit{Act} stage the proposal will notify to commanders and linked cyber defence experts the KDM findings, these will be persistently recorded at the IR, and if required, re-calibration indications may be communicated to cyber sensors and external data sources. The following describes the implemented procedure step-by-step:  

\begin{enumerate}
\item Monitorization. Aligned with the Perception layer of the Endsley's model, these preliminary steps initiates the KCT discovery loop by continuous monitoring the operational environment, including network, mission and the status related indicators of the deployed assets. This stage also includes the reception of feeds from external data sources, among them CTI, and information from NOCs, SOCs etc. and other facilities linked to the planned/ongoing mission.

\item Asset Discovery. During this step the ADM performs an active scan of the operational network to discover the assets in the network. Once all machines have been discovered, a port scanning is initiated not only to discover the service but also to determine the service version and even operating system. Applications as Nmap, Nessus or OpenVAS would be suitable to implement this scanning, moreover they are capable to retrieve assets' Common Platform Enumeration (CPE) intel supported by the external intelligence collected at Monitorization stage. Once the CPE is retrieved, a NIST National Vulnerability Database (NVD NIST) allows to request the CVEs and CVSS scores associated. This request is arranged by VDAM with the information extracted from the network scanning. At the end if the stage and from the perceived information, new knowledge about the operational assets and their vulnerabilities and scores is inferred and stored in IR.

\item Mission/Tasks Configuration. At this phase the cyber assets have been discovered and scored regarding their published vulnerabilities. In case some asset could not be identified automatically, the MTCM allows to manually insert assets into the system according to its CPE name, WFN (Well-Formed CPE Name) \cite{ares32}. The MTCM main functionality is to let a commander build the mission topology within inter-dependencies, so it is possible to define the operational tasks, its severity in the mission-centric context, and its inter-task relations and dependency degrees. It is also needed to specify the assets used per mission task and CoA, including the dependency degrees on them. Finally, in order to propagate intermediate computations for KCT discovery, inter-assets dependencies shall be provided.

\item Traffic Analysis. Based on the feeds of the Monitoring phase, the TAM analyzes network operational traffic supported by CTI towards computing a traffic score regarding availability, confidentiality and integrity measures.

\item KCT Discovery. In this step the KDM performs computations using task severity, task dependencies and asset dependencies, in order to calculate the criticality of the assets, for which the traffic score and linked vulnerability scores are considered. A deeper description of this calculation is described in the next section.

\item Internal Repository. Beyond the locally stored data that enables the calculations on the different proposed modules, at this stage the discovered assets and their criticality based on the mission context are permanently stored by the IR.

\item Notification. This phase includes the functionalities required to communicate the acquired cyber situational awareness on the deployed assets and their degree of criticality. This facilitates the decision and planning of related CoAs (if required), their inclusion on the Common Operational Picture (COP), the notification of the identified threats to the users, and if corresponds, their communication to external collaborators, stakeholders, authorities (local and international cybersecurity agencies, CSIRTS, Law Enforcement Agencies (LEAs), etc.) in compliance with the assumed common cyber defence policies.

\item Calibration. If supported by the deployed infrastructure, this stage groups the functionalities related with re-configuring the monitorization and comprehension capabilities towards verifying, evidence collection and/or presenting a more detailed analysis of the explored CIS terrains. For example, if a novel CIS asset is discovered but the consumed CTI does not provide much detail, the calibration functionalities may demand to query additional information on the external data sources. 
\end{enumerate}

\subsection{KCT Scoring and decision-making}
The conventional analysis of kinetic terrains tends to rely on the stages summarized by the KOCOA mnemonic acronym: Key Terrain/Decisive Terrain; Observation and Fields of Fire; Concealment and Cover; Obstacles; and Avenues of Approach/Withdrawal \cite{ref-KOCOA}; which shall converge towards describing a mission-centric network map \cite{newTDoS94}. In analogy with the observation, concealment, recognition of obstacles and avenues of approach/withdrawal (e.g., firewalls and port blocks); the existing advanced on cybersecurity have the potential of providing insight about the field (network and linked CIS asset), avenues of approach (by identifying nodes and links, which connect endpoints to specific sites), their vulnerabilities and dependent risks sores, and the capability of cyber commands on enforcing effective countermeasures. According to widely adopted procedures, as is the adoption of the CVSS scoring system, the raw CIS-level analysis has into account features like the adversarial skills required for weaponizing each vulnerability, the supportive tools that the attackers require, potential collateral damage (vertical/horizontal hybrid propagation), likelihood of an attack to success, or CIS-level valuation of the victim capabilities \cite{ares33}. The Table \ref{traceability-table} highlights the dimensions approached by the proposed methodology and the cyberspace factors to be considered for a mission-centric KOCOA terrain analysis.

\begin{table}[h!]
\centering
    \small
    \begin{tabular}{|m{2.1cm}|m{8cm}|m{5cm}|}
    \hline
    \textbf{Cyberspace Factor} & \textbf{KCT assessment considerations} & \textbf{Examples} \\ \hline
    \textbf{K}ey Terrain and Decisive   Terrain & The cyber assets   which MACS exceed the Mission Threshold (MTH) are considered mission KCTs. Decisive   terrains are KCTs which MACSs reaches their maximum value, so if jeopardized   the mission fails. & Critical Lines of Communications   (LOCs), tactical cloudlets, cyber command and control services, etc. They may   dynamically change as the mission evolves. \\ \hline
    \textbf{O}bservation and   Fields of Fire & The TBS is computed   in the TAM by analyzing the cyber field on which a planned/ongoing mission   occurs (network traffic), inferred from a CIS-level topology of the   operational environment. & Hosts, networks, services, cyber personas, etc. and linked cyber assets that may be   jeopardized in terms of confidentiality, integrity or availability. \\ \hline
    \textbf{C}oncealment and Cover & The TBS considers the   confidentiality, integrity and availability dimensions that describe the   field capabilities for addressing a secure use of the network. & Cyber bulletproof   services, anonymization, cryptography, of security policies, intrusion   prevention systems, etc. \\ \hline
    \textbf{O}bstacles & The calculation of VBS   relies on vulnerability metrics that have into account barriers like the   attack complexity by the analyzed vectors based on the deployed securization,   degree of user interaction or privileges required (see CVSS base metrics \cite{ares33}) & Black lists,   sandboxes, access control systems, firewalls, blockers,  recovery services, etc. Any technological, organizational or human thread prevention. \\ \hline
    \textbf{A}venues of Approach and Withdrawal & The VBS assesses   each cyber terrain by taking into account the qualities intrinsic to the   asset vulnerabilities, evolution and how each potential attack surface may be   weaponized by the attacker. & Vulnerabilities, open   network regions, gateways. DNS services, physical interfaces, network slices,   multitenancy, etc. \\ \hline
    \textbf{Mission}-centric vertical propagation & The ATAS provides   the final dependency degree between each asset and a given task. Based on   this, the TSAS provides information about how important is a task based on   commanders' predefined task severity in addition with the propagated severity   of its dependent tasks weighted by the dependency degree paths. & Dependencies between   cyber assets (services, communication channels, databases, etc.) regarding   mission tasks and their impact (delays, loss of surprise factor, reduced   decision-making agility, etc.). \\ \hline
    \end{tabular}
    \caption{\label{traceability-table}KCT discovery and assessment factors}
\end{table}

But as remarked in \cite{ref-hilltop}, \textit{"a critical omission in previous research efforts is the failure to tie
key terrain to objectives or missions"}, which need to create a convergence system withing the role of each studied assets, the aforementioned described of the cyber field, and the capability of the attacker on vertically propagate threats from the CIS asset to tactical, operational and strategic dimensions. In order to close this state-of-the-art gap, the following describes a global procedure for calculating the criticality of the mission terrains based on the operational network traffic features, linked vulnerabilities and dependent mission objectives and tasks. This is addressed thorough calculation of each asset linked Task Asset Criticality Score (TACS) and decisions made on them.

\subsubsection{Task Asset Criticality of mission assets}
Before establishing whether an asset is a KCT or not, they ought to be scored. The score system proposed is formed by four main calculations: Traffic Base Score (TBS), Vulnerability Base Score (VBS), Asset-Task Aggregated Score (ATAS), and Task Severity Aggregated Score (TSAS). These contributions make possible to define the equation that indicates each asset criticality on a task, based on not only technical characteristics but also strongly influenced over mission-oriented aspects. It is referred to as Task Asset Criticality Score (TACS), which is expressed as follows:

\begin{equation}
\label{eq:tacs}
\begin{split}
TACS_{a,t}=TSAS_t\cdot(\dfrac{3}{5}\cdot\:ATAS_{a,t}+\dfrac{1}{5}\cdot\:TBS_a+\dfrac{1}{5}\cdot\:VBS_a)
\end{split}
\end{equation}

\noindent  where \textit{a} is the analyzed CIS operational asset and \textit{t} is a mission task that depends on it. Note that for all of the metrics, the possible range values are constrained between 0 and 1. Thus, the Task Asset Criticality Score (TACS) is a normalized score between 0 and 1 to assess the assets criticality. The next subsections describe in detail each of the scores computed for composing the Task Asset Criticality Score (TACS).

\subsubsection{Traffic Base Score}
The considered Traffic Base Score (TBS) is computed in the TAM by analyzing the cyber field on which a planned/ongoing mission occurs: the network traffic. This may be served by well-known techniques like packet analysis, pattern detection or events triggered from an IDS or SIEM. The TBS approaches the challenge of summarizing the network status as a cyber field based on its confidentiality, integrity and availability during the supported operation, being the TBS calculated as described below:

\begin{equation}
TBS_a=\dfrac{1}{3}\cdot Avail_a+\dfrac{1}{3}\cdot Conf_a+\dfrac{1}{3}\cdot Int_a
\end{equation}

\noindent where \textit{a} is the CIST asset for which TBS is calculated, and the confidentiality, integrity and availability dimensions describe its capabilities for addressing a secure use of the network. Note that in the context of the conducted experimentation, and since TBS may present a diverse scope of scoring procedures (as it can be calculated from diverse sources of data), the TBS computation has been implemented in a simple but easily understandable way: by inferring the risk dimensions from packet inspections of the incoming/ongoing traffic on the analyzed CIS operational asset

In this context, the \textbf{Availability} (\textit{Avail}) metric aims on inferring the CIS operational asset needs for availability within the mission. The Availability computation is simplified by calculating the percentage of traffic and asset that are being used during the mission. This traffic is identified by observing the tuple integrated by the incoming/ongoing IP addresses and ports. They serve for tracking the packet-based traffic associated to the different network connections (identified by IP Address-port tuple) between the different assets through the operational network and allows to estimate a metric related to its availability needs. With experimental purpose, a sigmoid function has been proposed, which is applied to the traffic volume in order to spread near low values for better differentiation. A ranking-reward process has been applied in order to give more importance to the assets that are being more used in the network. Accordingly:

\begin{equation}
Avail_a=\dfrac{3}{5}\cdot Sigmoid+\dfrac{2}{5}\cdot RR
\end{equation}
for \textit{Sigmoid}:
\begin{equation}
Sigmoid=\dfrac{1}{1+100\cdot e^{-0.1x}}
\end{equation}
where \textit{x} is the percentage of traffic. For the \textit{Ranking Reward (RR)}:
\begin{equation}
RR=Sigmoid+(1-\dfrac{Rank}{Connections})-Sigmoid\cdot(1-\dfrac{Rank}{Connections})
\end{equation}

\noindent where \textit{Rank} is the position of the connection in the most-used connection ranking and \textit{Connections} is the total number of connections that are present in the network.

On the other hand, \textbf{Confidentiality} (\textit{Conf}) serves as flag that indicates the degree in which, whether or not, protocols or mechanisms for assuring confidentiality are implemented in the cyber communications of the analyzed CIS operational asset. In order to conduct a simple and easily understandable experimental implementation, confidentiality has been calculated based on the presence of TLS protocol in the communication while taking into account whether or not the communication uses a Cipher Suite stated as \textit{confidentiality secured} \cite{ares33}. Based on this, the confidentiality is expressed as follows:

\begin{equation}
Conf_a=\dfrac{3}{5}\cdot Confidentiality\;TLS\; set+\dfrac{2}{5}\cdot Confidentiality\;Secured
\end{equation}

\noindent where \textit{TLS set} is a flag that indicates if the connection is using TLS connection and \textit{Confidentiality Secured} is a flag activated if the TLS protocol is using a \textit{confidentiality secured} Cipher Suite.

Analogously, the network traffic \textbf{Integrity} (\textit{Int}) indicates the degree in which the same mechanisms assure integrity, so the presence of TLS integrity flags and a Cipher Suite \textit{integrity secured} will support the following calculation:

\begin{equation}
Int_a=\dfrac{3}{5}\cdot Integrity\;TLS\;set+\dfrac{2}{5}\cdot Integrity\;Secured
\end{equation}

\subsubsection{Vulnerability Base Score}
The Vulnerability Base Score (VBS) is related with the kinetic KOCOA analysis on concealment and cover, obstacles, and avenues of the enemy approach. In the grounds of the CVSS scoring system (base metrics), the VBS analyzes each cyber terrain by taking into account the qualities intrinsic to the asset vulnerabilities, their expected evolution over lifetime, and how each potential attack surface may be weaponized by the attacker based on elements like complexity or potential access vectors. During the conducted experimentation the VBS has been directly retrieved from VDAM through requests to the NVD NIST CVE web service. In this repository each asset (characterized by a CPE entity) is related to a list of vulnerabilities (CVE) and their scores (CVSS). In this approach it is taken into account the maximum value of each of the CVSS Base Score value related to the asset identified as follows:

\begin{equation}
\label{eq: vbs}
VBS_a=\dfrac{1}{10}Max\{CVSS\:Base\:Score_{a,1},...,CVSS\:Base\:Score_{a,n}\}
\end{equation}

\noindent where \textit{a} is a considered cyber asset and \textit{n} the number of CVSS associated to the asset \textit{a}.

\begin{figure*}[t!]
  \centering
  \includegraphics[width=0.8\linewidth]{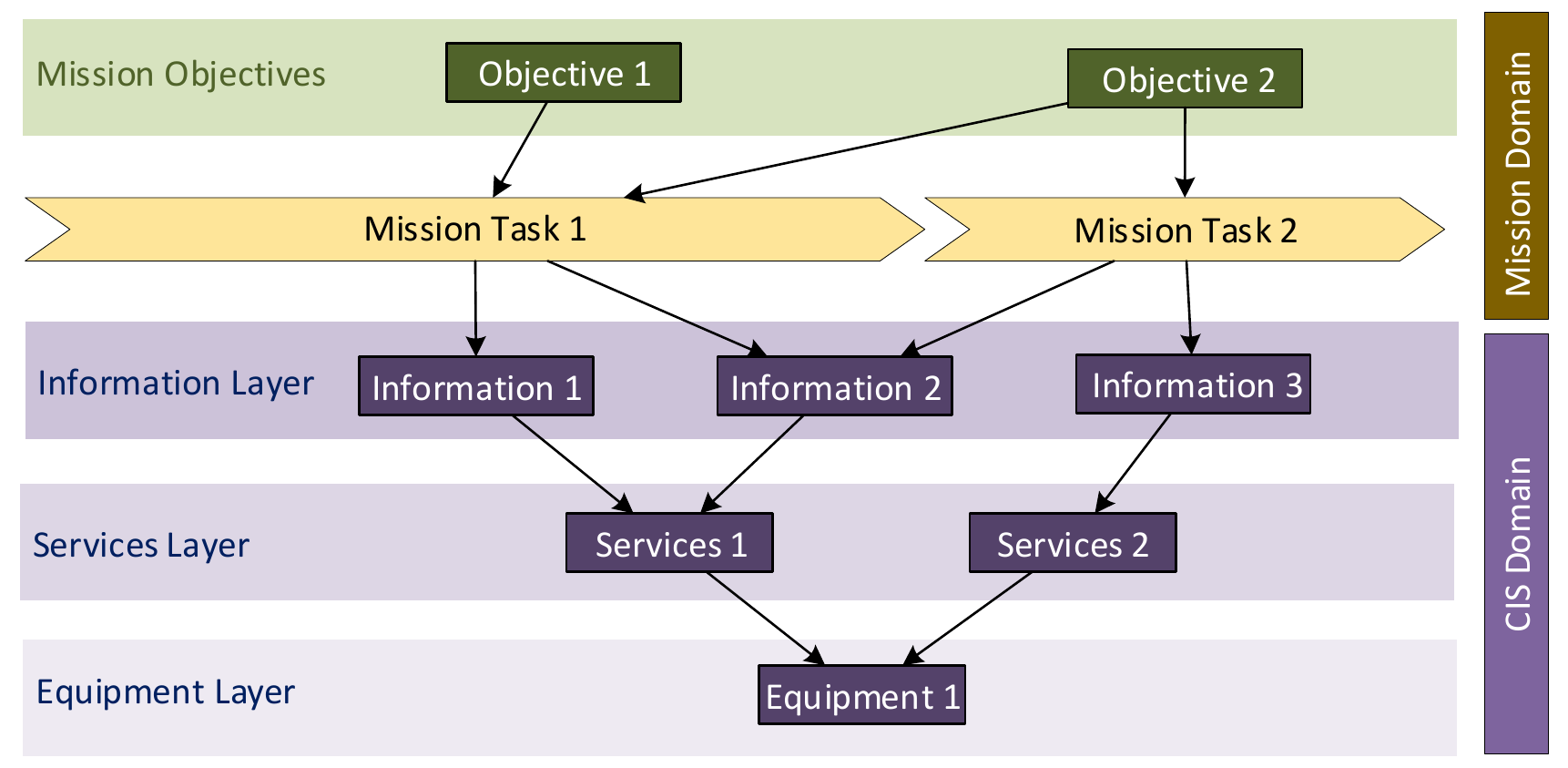}
  \caption{  \label{figdepe}Top-level mission-centric dependecies.}
\end{figure*}

\subsubsection{Asset-Task Aggregated Score}
The Asset-Task Aggregated Score (ATAS) provides the final dependency degree between each asset and a given task, being computed at the KDM. ATAS takes into account the analyzed task's dependency for each asset on which its proper execution relies on, as well as the linked assets' inter-dependency paths. This dependency path is computed as suggested in Book III of Magerit methodology \cite{ares10}, but bearing in mind the mission-centric aspects (relationship between tasks and goal). In compliant with Magerit's guidelines, the resulting dependency trees shall consider information and services at the top, where the KDM includes two additional layers: Tasks and Objectives. The resulting hierarchy is illustrated in Figure. \ref{figdepe}, which may be varied by cyber defence experts if the circumstances demand it. Alternatively, and if the top-level dependencies tree shall consider hybrid aspects (e.g., cyber persons, social implications, etc.) \cite{new-hybridDamage}, additional levels may be considered. Note that in order to facilitate the understanding of the proposal, these hybrid elements were not considered during its implementation and empirical validations. However, it is important to extend these graphs when strategically and operationally thinking on the cyberspace, which is discussed in wider detail at Section 6. 
The asset-task dependencies may be direct (e.g., in Figure. \ref{figdepe}, \textit{Mission Task 1}, $\rightarrow$ \textit{Information 1}); which ocurr were there are not transitive assets to be taken into account; and independent (e.g., in Figure. \ref{figdepe}, \textit{Mission Task 1} $\Rightarrow$) \textit{Equipment 1}), where the degree of dependency is propagated. The resulting Asset-Task Score (ATS) is propagated based on each dependent value and inter-dependencies, being expressed as follows:

\begin{equation}
\label{eq: ATAS}
ATAS (t,x) = 1-\prod_i^{i=p}({1-degree(t \Rightarrow x)}_i) 
\end{equation}

\noindent where $x$ is a mission asset, $t$ refers to the particular task on which ATAS is calculated, and $n$ is the total number of dependency path that connect $t$ with $x$. In coherency with Magerit, the degree of dependency is modelled as a continuum between 0.0 (independent assets) and 1.0 (fully dependent assets). This value is provided by the ADM when the assets are discovered.

An example of ATAS calculation can be described from the example presented at Fig. \ref{figdepe}. Based on this, let the following paths:

\begin{equation}
\label{eq: ATAS-ex1}
degree(Mission\;Task\;1 \rightarrow Information\;1 \rightarrow Services\;1) = 0.3
\end{equation}

\begin{equation}
\label{eq: ATAS-ex2}
degree(Mission\;Task\;1 \rightarrow Information\;2 \rightarrow Services\;1) = 0.5
\end{equation}

\noindent it is possible to conclude that:

\begin{equation}
\label{eq: ATAS-ex}
ATAS(Mission\;Task\;1, Services\;1)= 1 - (1 - 0.3)\times(1 - 0.5)
\end{equation}

\noindent since these are the only dependency paths that define the dependencies between $Mission\;Task\;1$ and $Services\;1$.

\subsubsection{Task Severity Aggregated Score}
The purpose of the Task Severity Aggregated Score (TSAS) is to evaluate the total severity associated with a task, which will vary depending on the tasks that depend on it according to the degrees of mission-level dependencies identified by the mission manager. TSAS presents values between 0 and 1, being calculated similarly to ATAS at the KDM. This metric provides information about how important is a task based on commanders' predefined task severity in addition with the propagated severity of its dependent tasks weighted by the dependency degree paths. Let the mission task $t$, the TSAS correlates its cumulative severity against each other $y$ task ($CS_{t,y}$), which is expressed as follows:

\begin{equation}
\label{eq:TSAS-SA}
CS_{t,y} = severity(y)\times degree(t\Rightarrow y)
\end{equation}

\noindent being $TSAS_t$ calculated as described below:

\begin{equation}
\label{eq:TSAS-CS}
TSAS_t = 1 - (1-severity(t)) \prod_i^{i=q} (1 - CS_{t,i})
\end{equation}

\noindent where $q$ is the total number of tasks on which $t$ relies on, hence $CS_{t,i}$ expressing the cumulative severity regarding each of them.

\subsubsection{Task/Mission Asset Criticality Score}
As stated at the beginning of this section, the four metrics previously defined, are used to compute the assets' criticality for each task (TACS) (Equation~\ref{eq:tacs}). Once TACS is calculated for each task and asset, a basic Task Threshold (TTH) is defined in order to set whether an asset is critical and consequently a KCT for that task. The assets which TACS is above that Task Threshold is identified as a task KCT.

\begin{equation}
\begin{split}
\label{eq:tth}
TTH_t = Avg\{TACS_{1,t},...,TACS_{A,t}\}+\\
+\:k\cdot StdDev\{TACS_{1,j},...,TACS_{A,t}\}
\end{split}
\end{equation}

\noindent where \textit{t} is the task to calculate the threshold, \textit{A} the total number of assets involved in the task \textit{t}, and \textit{k} is a tuning parameter (between 0 and 1) used to adjust the sensitivity of the KCT identification. \textit{k} value should be selected in the implementation depending on the commander's chosen approach: optimistic (values around 1), medium (values around 0.5) or pessimistic (values around 0). Selecting a medium approach it is more appropriate since it might avoid possible  false-negatives and false-positives.

From a global mission perspective, it is also suggested in this methodology to calculate the mission KCTs to have an overview of critical assets in order to help commanders or decision-makers to select possible CoAs. Thus, a Mission Threshold (MTH) is defined in a similar way to TTH:

\begin{equation}
\begin{split}
\label{eq:mth}
MTH = Avg\{TTH_{1},...,TTH_{T}\}+\\
+\:k\cdot StdDev\{TTH_{1},...,TTH_{T}\}
\end{split}
\end{equation}

\noindent where \textit{T} is the total number of tasks in the mission and \textit{k} a tuning parameter with the same objective as stated for the Task Threshold.

One asset has associated as much Task Asset Criticality Scores (TACS) as tasks using that asset. To calculate Mission Asset Criticality Score (MACS) for the asset it is needed to select the maximum value of the scores set. 

\begin{equation}
\begin{split}
\label{eq:macs}
MACS_a=Max\{TACS_1,...,TACS_M\}
\end{split}
\end{equation}

\noindent where \textit{M} is the total number of tasks in which an asset \textit{a} is used. Accordingly, in the context of the conducted implementations and experimentation, every asset which MACS exceed the Mission Threshold (MTH) is considered mission KCT. This criteria may be statically or adaptive modified based on the operational circumstances, and even tailored to pessimistic, average or optimistic scenarios. For example, in the TACS Equation \ref{eq:tacs} described at section 4.3.1, the weighs given to each element the assessment of Task-Active Dependencies is greater against other metrics  since it was assumed by design that the identification of KCT should respond to a large extent to criteria related to mission needs ($\dfrac{3}{5}$ at $ATAS_{a,t}$), resulting in a less weighing for the more technical aspects of the mission ($TBS_a$ and $VBS_a$), but without neglecting the latter. Hence alternative $TACS_{a,t}$ computations may be built as

\begin{equation}
\label{eq:tacsD}
\begin{split}
TACS_{a,t}= TSAS_t \cdot(Mw \cdot \: ATAS_{a,t}+ Bw\cdot\:TBS_a+Tw\cdot\:VBS_a)
\end{split}
\end{equation}

\noindent where $Mw$ is the custom weight for $ATAS_{a,t}$, $Bw$ is the custom weight for $TBS_a$, and $Tw$ is the custom weight for $VBS_a$; so $Mw, Bw, Tw \in [0 \cdots 1]$.     

\section{Capability development and military thinking}
Beyond the technical effectiveness achievable by the proposed mission-centric CKT analytics, their evolution and end-user consideration are constrained by intrinsic tactical, operational and strategical challenges, returning in benefits and drawbacks that shall be explored and when possible, mitigated. In order to contribute to this exercise while facilitating a better understanding of the topics and novelties described above, the following presents a challenge-directed KCT capability building analysis on the presented solutions. This is complemented with a discussion on their potential relevance in terms of military offensive and defensive thinking, the latter aiming on facilitating the identification of future doctrine-level implications and disruption.

\subsection{KCT Capability Building Analysis}
As widely adopted by defence practitioners, the DOTMLPF-I construct presents the most common measures of the status of a military capability thorough the analysis of the Doctrine, Organization, Training, Materiel, Logistics, Personnel, Facilities, and Interoperability of the targeted solution \cite{dompfi}. They are discussed as follows:

\subsubsection{Doctrine} 
As pointed out in \cite{ref-ThreeSwords}, there is a huge gap concerning the cyber defence doctrine, that ranges from cyber related definitions and taxonomies, to Rules of Engagement (RoE), and \textit{just ad/in/post bellum} considerations. Despite of this, the concept Key Cyber Terrain is analogous to its preliminary meaning at kinetic operations, which facilitates the understating of their commonalities. However, the NATO Joint Publication for Cyberspace Operations \cite{newTDoS94} points out a pair of major differentiating aspects that may be complemented by other insights on related doctrine, as is the case of the fact that an adversary may occupy the same terrain or use the same process in cyberspace, potentially without knowing of the other’s presence; or the fact that each cyber terrain has a virtual component, identified at logical network or even cyber-persona layers. Overall, joint mission planning shall foresee the accessibility to key terrain in blue, gray, and red cyberspace for each plan. Similarly to \cite{newDoctrine-offensive}, KCTs became decisive terrains when their seizure and retention are mandatory for successful mission accomplishment.

\subsubsection{Organization} As common in all cyber defence application, the proper discovery and assessment of cyber terrains require at least to establish cross-functional entities and governance procedures for harnessing related expertise and information. When supporting joint or combined operations, KCT discovery services shall address difficulties inherent in monitoring and collecting information from CIS environments with multiple security domains and policies, where related communication and informational procedures must be compliant with common harmonization, regulation and standardization procedures. Based on this, the organization constitutes one of the most critical aspects of a success KCT analysis, where additional factors may be considered, among them  the potential need on involving specific technical skills, the independence of the KCT analytical team from other resources, the fact that some CIS assets on which the mission may directly or indirectly rely on shall be explored, or the need for assistance by external actors (e.g., local CERTs, NOCs, SOCs, etc.) \cite{ref-capabilities-organization} . When closer to the strategic level, the need for a stronger organizational ability increases. This tends to decrease when approaching to operational, tactical and technological levels, the latter usually increasingly skills and more complex technological enablers. 

\subsubsection{Training}
Against the growing global need for cybersecurity and cyber defence experts, there is an exacerbated shortage of professionals with the Knowledge, Skills and Abilities (KSA) required for conducting military operations on the cyberspace, where the proposed KCT is expected to contribute. Nowadays, cyber preparedness is mostly: 1) traditional training and lifelong learning approaches, either using fully theoretical or limited hands-on adversarial approaches, that fail to equip professionals with real-life KSA to understand a complex cyber operational picture; 2) limited access to advanced evidence-based cost and impact analyses, resulting in questionable selection of cybersecurity measures from the cost/level of protection; and 3) limited coverage of cyber education and training able to transform mission-centric KCT user needs into tailored exercises and scenarios \cite{ref-capabilities-nice,FURNELL20206} . A proper KCT analysis also suggests the need for training professionals capable of bridging the gap between the technological, human, legal and commercial aspects of the cyberspace, which depict the side context on which each cyber terrain operate \cite{newCSA1}.

\subsubsection{Materiel}
The material required for the proper adoption of the proposed methodology suggests the need for deploying specialized third-party solutions, which are expected to be matured thorough years of development and adoption by cybersecurity practitioners. For example, one of the main challenges for the cyber defence material is to enable capabilities for monitoring and discovering the different indicators (periodic reports, alerts, events, etc.) from both physical and virtual changes on the CIS environment \cite{ZHOU20189}. They among others may bring functionalities for discovering the incorporation of new assets on the operational environment while identifying their vertical/horizontal dependencies, recognize removal or modification events on them, and/or notify any other situation that may be relevant for KCT analysis like bandwidth status, presence of unused resources or requests for special configurations. On the other hand, the required material shall be able to detect threats and outliers and infer mission-centric risks on the operational environment, thus demanding the connectivity with IDSs, SIEMS, SOCs, CERTs, mission planners and any other related data source; the latter also including the support to the connectivity with external data sources (API managers, Kubernetes, federated databases, etc.) \cite{ZARPELAO201725}. Another example is the possibility on relying on premade materiel able to facilitate the human understanding of the deployed assets while visually highlighting KCT and their dependencies \cite{refMe-Sal1}. Although this paper presents the KCT discovery and assessment proposal as abstracted as possible of its dependent technological background, it is assumed that the aforementioned cyber sensors, aggregators and information exchange functionalities are available and fully operational on the theatre of operations, so it can receive the adequate inputs on which its functionalities operate.

\subsubsection{Leadership} At KCT assessment, the leadership factor significantly differs from the decisions and actuation skills inherent to orchestrating incidence response teams at operational level. On the contrary, KCT assessment leaders shall be able to coordinate the personnel involved in the management of the information feeds, their analysis, and the transference and integration of the acquired knowledge to a common operational picture. As usual in cyber defence, leaders involved in KCT assessment must have a strong background in security and be recognized for
its technical expertise \cite{BUCHLER2018114} .

\subsubsection{Personnel} The professionals involved in KCT assessment shall be part of well-trained teams deployed as close as possible to the CIS environment on which the cyber operation occurs. As indicated at Organization, Training and Personnel, they require the capabilities of understanding and adapting to the dynamic nature of the cyberspace, as well as the multiplicity of factors that may impact on the data flows to be processed, including their information security and ethical implications \cite{ref-capabilities-doctrine}. Since the cooperation between military cyber experts and civilians is mandatory, the KCT assessment involved personnel shall be familiar with the information exchange and good practices between cross-sectorial teams, which as commented may include Computer Emergency Response Teams (CERTs), NOC engineers and the military staff involved in the ongoing/planned related operations. Given the well-known shortage of cybersecurity suitably qualified staff and the fast evolution of the sector, KCT assessment involved personnel shall be continuously trained and specialized on the raising CIS enablers, information exchange procedures and adversarial tactics for thwarting the network monitoring and analysis actions. 

\subsubsection{Facilities} The facilities that hold the  physical and virtual infrastructure for KCT discovery and assessment must ensure a proper safety and security condition, hence embracing confidentiality, integrity and availability against both kinetic situations (electrical issues, coverage against natural threats, etc.) up to cyber situations (miss configurations, cyber attacks, data leaks, etc.). These facilitates may be external (e.g., assistant SOCs, NOCs, etc.) or owned (e.g., military CERTs, tactical bare-metal orchestrating cloudlets, antennas, etc.). In analogy with other cyber defence capability, the man-made nature of the cyberspace and its well-known reachability and unpredictability features depict an trend towards adopting distributed infrastructure able to distribute C3 services thorough the operational edge while increasing their resiliency against healing, optimization and protection issues \cite{doi:10.1080/14702436.2015.1108108} .

\subsubsection{Interoperability}
the assessment of cyber terrains demands doctrinal and technical interoperability, which intensifies when operating at cross-domain networks during joint, combined or military-civilian cooperative scenarios \cite{ref-capabilities-facilities}. In this context, cyber intelligence and information sharing is essential to infer a common cyber operational picture, which will reveal the deployed CIS assets dependencies, status and mission-centric effectiveness required for KCT analytics. From the doctrinal and operational perspective, and as preliminary introduced in Doctrine, Leadership and Personnel, the actors involved in KCT assessment must harmonize their  \textit{modus operandi} towards conducting agile, secure and cross-domain singularity compliance joint functions.  
On the other hand, the technological interoperability demands standardized, harmonized or certificated communication mechanisms (communication platforms, semantics, data models, protocols, STANAGS, etc.).

\subsection{KCT assesment at Offensive thinking}
According to the JP 3-12 \cite{newTDoS94}, offensive cyberspace operations are missions that intend to project power in and through foreign cyberspace, which literally \textit{may exclusively target adversary cyberspace functions or create first-order effects in cyberspace to initiate carefully controlled cascading effects into the physical domains}. On these grounds, targeting KCTs and/or decisive terrains is prioritized, as long as the cost-benefit results of the offensive actions allow it. This analysis requires their previous identification and assessment in analogy with the method described at previous sections, but transposing the ally mission-centric context to the adversarial regions on the cyberspace (i.e. the enemy's terrains). Based on this, some of the scoring metrics presented above have to be slightly reframed as summarized in Table \ref{traceability-Off}, where the enemy's networks and CIS assets are enumerated and their dependencies regarding the ally offensive steps substitute the variables described at Section 4.

Based on \cite{newDoctrine-offensive}, beyond the classical goals in offensive missions (defeat, destroy, or gain control on tactical objectives), commanders may target adversarial KCT for: 1) secure decisive ally kinetic or cyber terrains; 2) deprive the enemy of the digital resources provided by the KCT; 3) gain information on assets linked to the KCT; 4) Deceive and divert a kinetic or cyber enemy force supplied by the KCT; 5) Fix an enemy force in position; 6) disrupt an enemy attack by causing the triggering of contingency CoAs or power reduction; and 7) set the conditions for successful future operations.

\begin{table}[ht!]
\centering
\small
\begin{tabular}{|m{2.1cm}|m{12cm}|}
\hline
    \textbf{KCT Score} & \textbf{Offensive reframing} \\ \hline
    $TACS_{a,t}$ & The Task Asset Criticality Score (TACS) is computed based on the offensive mission task $t$ that targets the $a$ CIS asset owned by the adversarial. \\ \hline
    $TBS_a$ & The Traffic Base Score (TBS) is computed based on the capabilities   of the adversarial for addressing a secure use of its network asset $a$. \\ \hline
    $VBS_a$ & The Vulnerability Base Score (VBS) analyzes the CIS asset $a$ owned by the adversary by taking   into account the qualities intrinsic to the asset vulnerabilities, their   expected evolution over lifetime, and how each potential attack surface may   be weaponized by the ally forces. \\ \hline
    $ATAS (t,x)$ & The Asset-Task Aggregated Score (ATAS) shall provide the final dependency degree between the   targeted adversarial asset $x$ and a given $t$ ally offensive task, where a   0.0 score indicates non-relevance a 1.0 score reveals that $x$ shall be   jeopardized for $t$ success. \\ \hline
    $TSAS_t$ & The Task Severity Aggregated Score (TSAS) shall compute the score that quantifies the relevance   of an ally offensive task $t$ based on the propagated severity degrees of the   dependent tasks on the same offensive mission. \\ \hline
    $MACS_a$ & The Mission Asset Criticality Score (MACS)  presents the   most critical offensive task on which the enemy CIS asset $a$ is targeted. \\ \hline
\end{tabular}
    \caption{\label{traceability-Off}KCT scoring at offensive operations}
\end{table}

\subsection{KCT assesment at Defensive thinking}
Defensive cyberspace operations tend to focus on safeguarding the ongoing/planned mission assurance by impeding and recovering from adversarial actuations on the cyberspace \cite{newTDoS94}, which may be approached by enforcing internal and/or external cyber defensive measures. In both cases, the proper discovery and assessment of KCT entails an essential success factor. At internal operations, KCTs are ally CIS assets within the protected perimeter, so the proposed mission-centric analytics explore the significance of each ally asset regarding the ally mission tasks which proper execution relies on them. The Table \ref{traceability-DEF} summarizes the adoption of the KCT scoring systems described at Section 4 when applied to internal defensive measures.

\begin{table}[ht!]
\centering
\small
\begin{tabular}{|m{2.1cm}|m{12cm}|}
\hline
    \textbf{KCT Score} & \textbf{Defensive reframing} \\ \hline
    $TACS_{a,t}$ & The Task Asset Criticality Score (TACS) is computed based on the defensive mission task $t$ that aims on protecting or reinforcing the $a$ CIS asset, which may be within the ally network or external. \\ \hline
    $TBS_a$ & The Traffic Base Score (TBS) is computed based on the deployed capabilities for safeguarding the usability of the protected network asset $a$, \\ \hline
    $VBS_a$ & The Vulnerability Base Score (VBS) analyzes the protected or reinforced CIS asset $a$ by taking   into account the qualities intrinsic to the asset vulnerabilities, their   expected evolution over lifetime, and how each potential attack surface may be exploited by the attacker. \\ \hline
    $ATAS (t,x)$ & The Asset-Task Aggregated Score (ATAS) shall provide the final dependency degree between the protected or reinforced asset $x$ and the $t$ defensive task, where a 0.0 score non-dependence a 1.0 score reveals that $x$ shall be completely protected $t$ success. \\ \hline
    $TSAS_t$ & The Task Severity Aggregated Score (TSAS) shall compute the score that quantifies the relevance of the $t$ defensive task based on the propagated severity degrees of the   dependent tasks on the same defensive mission. \\ \hline
    $MACS_a$ & The Mission Asset Criticality Score (MACS)  presents the   most critical defensive task on which the CIS asset $a$ shall be protected or reinforced. \\ \hline
\end{tabular}
    \caption{\label{traceability-DEF}KCT scoring at defensive operations}
\end{table}

On the other hand, external cyber defensive maneuvers are conducted out of ally network, in cyberspace regions owned by neutral or hostile actors. These responses require very precise objectives, planning and Rules of Engagement (ROE), since the hybrid propagation of the actors may be confused by just ad/in bellum by neutral actor and/or difficult further ally steps. When the external maneuvers entail to weaken the attacker power, KCTs may be assessed as considered at offensive cyber maneuvers (see Section 5.2) so the indications summarized in Table \ref{traceability-Off} should be considered. On the opposite, when the external cyber maneuver aims on reinforcing the security perimeter or deploying attack prevention countermeasures, the external terrains can be computed as own, so the indications in Table \ref{traceability-DEF} can be adopted individually, or together with those analyzed at external maneuvers \cite{newTDoS94}.

\section{Case of study: Offensive Cyberspace Operation (OCO)}
This section describes an illustrative use case that applies the proposed methodology in order to clarify how KCTs are identified for a mission. The use cases is focused in hypothetical asset discovery, mission topology definition, traffic analysis, and KCT discovery on real traffic traces. The scenario is developed to show both offensive and defensive thinking for the methodology validation approach.

\subsection{Mission description}
A military mission narrative is proposed below to set the context, objective, assets and tasks to be carried out during the mission. In the context of an espionage operation against an enemy state, a mission is defined to gather information on the next actions that the enemy state is going to carry out in the coming months. 
To this end, a mission is designed to infiltrate an enemy facilities where it will connect to the internal network to access and transfer information from its servers. Beforehand, a UAV flight is carried out to obtain photographs in order to choose the possible infiltration routes.

\subsection{Asset Discovery and VBS}
Let us assume that during a multi-domain military operation, an operational network scanning using ADM discovered six cyber assets: A1, A2, A3, A4, A5, A6. The VDAM has retrieved their CVSS Base Score \cite{ares33} to compute their VBS as summarized in Table~\ref{tab:vbs}.

\begin{table}[ht!]
\centering
  \begin{tabular}{|c|c|}
  \hline 
    \textbf{Cyber Asset}&\textbf{Vulnerability Base Score}\\\hline 
    
    A1 & 0.65\\ \hline 
    A2 & 0.95 \\ \hline 
    A3 & 0.33\\ \hline 
    A4 & 0.64\\ \hline 
    A5 & 0.26\\ \hline 
    A6 & 0.62\\ \hline 
\end{tabular}
  \caption{\label{tab:vbs} Vulnerability Base Scores (VBS)}
\end{table}

\subsection{Mission topology definition}
\begin{figure}[t]
  \centering
  \includegraphics[width=0.6\linewidth]{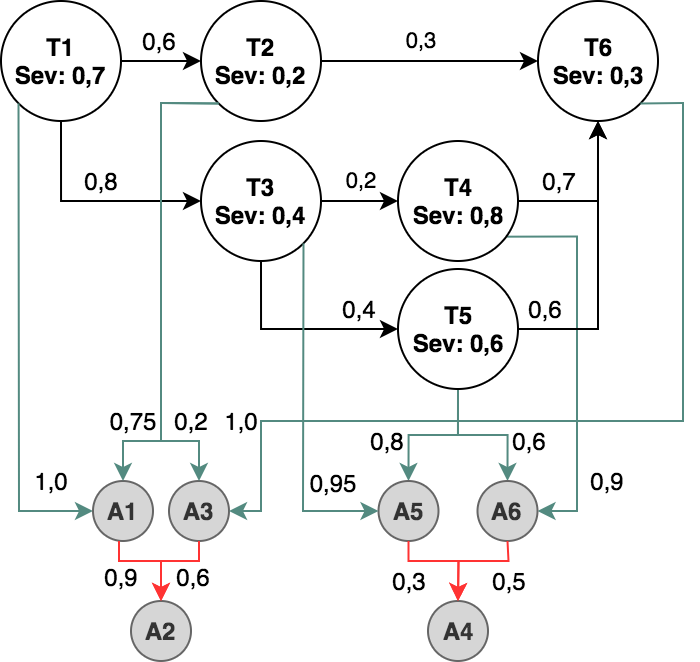}
  \caption{\label{fig3}Mission topology and inter-dependencies.}
\end{figure}

After the asset discovery stage, the commander may define the mission topology (see Fig. \ref{fig3}) as depicted in the MTCM. For the current mission six tasks with their associated severities and inter-task dependency degrees were identified. In addition, relationships among assets and mission tasks, and the cyber assets inter-dependencies shall be also defined.

\subsection{Traffic analysis}
A mission simulation considering both kinetic and cyber domains has been conducted in the grounds of Indra's Synthetic Mission Generator (ISMG) in order to analyze traffic-based characteristic as discussed in Section 4. On these results, the TAM module detects traffic involving A1 and A2 is using a secure TLS Cipher Suite whereas A5 and A6 are using a TLS non-secure cipher suite. A3 and A4 are not using ciphered protocols. In addition, traffic percentage is calculated to determine the availability requirements of the assets. Finally, with this information the assets traffic scores are calculated as illustrated in Table~\ref{tab:tbs}, by implementing the Equation \ref{eq:tbs}.

\begin{table}[ht!]
\centering
  \begin{tabular}{|c|c|}
     \hline
    \textbf{Cyber Asset}&\textbf{Traffic Base Score}\\
     \hline
    A1 & 0.589\\ \hline
    A2 & 0.494\\ \hline
    A3 & 0.103\\ \hline
    A4 & 0.331\\ \hline
    A5 & 0.256\\ \hline
    A6 & 0.471\\ \hline
\end{tabular}
  \caption{\label{tab:tbs} Traffic Base Scores (TBS)}
\end{table}

\subsection{KCT Discovery}
The last step to discover the KCTs is to calculate the Task/Mission Asset Criticality (TACS and MACS). This step requires previously computed Traffic (TBS) and Vulnerability (VBS) scores, as well as computing Asset-Task Aggregated Score (ATAS). By the use of these three metrics, the final criticality asset score for the task is computed weighted by the Task Severity Aggregated Score (TSAS).\\
Firstly, the Asset-Task Aggregated Score (ATAS) is calculated per asset and task (Table~\ref{tab:atas}). The outcome indicates how important the cyber asset is at that time snapshot, bearing in mind the cyber assets inter-dependencies.

\begin{table}[ht!]
\centering
  \begin{tabular}{|c|c|c|c|c|c|c|}
    \hline
    \textbf{Asset}&\textbf{T1}&\textbf{T2}&\textbf{T3}&\textbf{T4}&\textbf{T5}&\textbf{T6}\\
    \hline
    A1 & 1.0    & 0.75  & 0     & 0     & 0     & 0\\ \hline
    A2 & 0.9    & 0.714 & 0     & 0     & 0     & 0.6\\ \hline
    A3 & 0      & 0.2   & 0     & 0     & 0     & 1\\ \hline
    A4 & 0      & 0     & 0.285 & 0.45  & 0.468 & 0\\ \hline
    A5 & 0      & 0     & 0.95  & 0     & 0.8   & 0\\ \hline
    A6 & 0      & 0     & 0     & 0.9   & 0.6   & 0\\ \hline
\end{tabular}
\caption{\label{tab:atas} Asset-Task Aggregated Scores (ATAS)}
\end{table}

After that and taking into account the task severity inter-dependencies, the Task Severity Aggregated Score (TSAS) is computed as illustrated in Section 4.3.4 (see Table~\ref{tab:tsas}). 

\begin{table}[ht!]
\centering
  \begin{tabular}{|c|c|c|}
     \hline
    \textbf{Task}&\textbf{Severity}&\textbf{Task Severity Aggregated Score}\\
     \hline
    T1&0.7&0.889\\ \hline
    T2&0.2&0.272\\ \hline
    T3&0.4&0.633\\ \hline
    T4&0.8&0.842\\ \hline
    T5&0.6&0.672\\  \hline
    T6&0.3&0.300\\ \hline
\end{tabular}
  \caption{\label{tab:tsas} Task Severity Aggregated Scores (TSAS)}
\end{table}

As a result of the previous metrics calculations (TBS, VBS, ATAS and TSAS), it is possible to obtain the final criticality score of each asset involved on each task. For that, the Equation \ref{eq:tacs} takes into place to compute the Task Asset Criticality Score (TACS) (see Table~\ref{tab:tacs}). TACS is designed in order to strengthen the mission-centric perspective of the asset's criticality over asset's technical characteristics.
\\Once TACS is calculated for each task and asset, the Task Threshold (TTH) might be computed in order to decide whether or not an asset is KCT and thus is tagged as critical for the task success (see Table~\ref{tab:tacs}). The threshold is calculated as defined in Equation \ref{eq:tth}. In section 4, was mention the need of setting the value for the tuning parameter \textit{k} in order to adjust the sensitivity of the system to identify KCTs. In this implementation a medium approach is being performed (\textit{k = 0.5}) to avoid possible false-positives and false-negatives (Equation \ref{eq:tthimpl}).

\begin{equation}
\begin{split}
\label{eq:tthimpl}
TTH=Avg\{TACS,...TACS\}+\\+\dfrac{1}{2}\cdot StdDev\{TACS,...TACS\}
\end{split}
\end{equation}

\begin{table}[t!]
\centering
  \begin{tabular}{|c|c|c|c|c|c|c|}
    \hline
    \textbf{Asset}&\textbf{T1}&\textbf{T2}&\textbf{T3}&\textbf{T4}&\textbf{T5}&\textbf{T6}\\
    \hline
    A1 & 0.754  & 0.190     & 0     & 0     & 0     & 0\\\hline
    A2 & 0.737  & 0.195     & 0     & 0     & 0     & 0.195\\\hline
    A3 & 0      & 0.056     & 0     & 0     & 0     & 0.206\\\hline
    A4 & 0      & 0         & 0.231 & 0.391  & 0.319 & 0\\\hline
    A5 & 0      & 0         & 0.426 & 0.087  & 0.392   & 0\\\hline
    A6 & 0      & 0         & 0     & 0.510 & 0.389  & 0\\\hline
    
 \hline
  \emph{TTH} & 0.751         & 0.186 & 0.398  & 0.510 & 0.387 & 0.204
\end{tabular}
  \caption{\label{tab:tacs} Task Asset Criticality Scores}
\end{table}

Finally, from a mission perspective, the Mission Criticality Asset Score (MACS) is obtained from the maximum value of the asset for all tasks (see Table~\ref{tab:macs}) and the Mission Threshold (MTH) is computed in an analog way as followed for Task Threshold, using a medium approach (Equation \ref{eq:mthimpl}).

\begin{equation}
\label{eq:mthimpl}
\begin{split}
MTH=Avg\{TTH_1,...,TTH_N\}+\\+\dfrac{1}{2}\cdot StdDev\{TTH_1,...,TTH_N\}=0.511
\end{split}
\end{equation}

In view of the results in Table~\ref{tab:tacs} and Table~\ref{tab:macs}, it is possible to state that the KCTs for the Task 1 is A1; for Task 2, are A1 and A2; for Task 3 is A5; for Task 4 is A6; for Task 5 is A6 and for Task 6 is A3. \\From the global mission perspective, assets A1, A2 and A6 are identified as KCTs for the overall mission, which represent the most valuable terrains to be taking into consideration at C2 and decision-making actions.

\begin{table}[ht!]
\centering
  \begin{tabular}{|c|c|}
    \hline
    \textbf{Cyber Asset}&\textbf{MACS}\\
    \hline
    A1 & 0.754\\\hline
    A2 & 0.737\\\hline
    A3 & 0.206\\\hline
    A4 & 0.391\\\hline
    A5 & 0.426\\\hline
    A6 & 0.638\\\hline
\end{tabular}
  \caption{\label{tab:macs} Mission Asset Criticality Scores (MACSs)}
\end{table}

\section{Conclusion}
The identification of Key Cyber Terrains (KCT) is crucial at cyber-oriented missions' operation, since they allow commanders and decision-makers to notice what might be essential to preserve with the objective of achieving a mission success. An in-depth review of the state of the art revealed important gaps and challenges concerning the assessment of each cyber asset critically while assuming mission-centric conditions in dynamic operational contexts. As response, the conducted research introduced a novel approach for dynamically infer KCTs bearing in mind each mission task and overall goals. However, and despite the foreseen effectiveness of the solution, this paper describes the outcomes of an early research that assumed important limitations, most of them being expected to be explored as part of future works. On the other hand, emerging topics like adversarial machine learning may be studied as potential weapons able to thwart the cyber terrain scoring systems targeting to mislead decision-makers. Although the methodology is design purely for a military perspective it might be applied for a dual-use, not only for a military application, but also for civil or emergencies since the model is aware of tasks regardless their purpose. 

\section*{Disclaimer}
The contents reported in the paper reflect the opinions of the authors and do not necessarily reflect the opinions of the respective agencies, institutions or companies.
\section*{Acknowledgement}
\hfill \break
\begin{minipage}{0.3\linewidth}
    \includegraphics[width=0.8\textwidth]{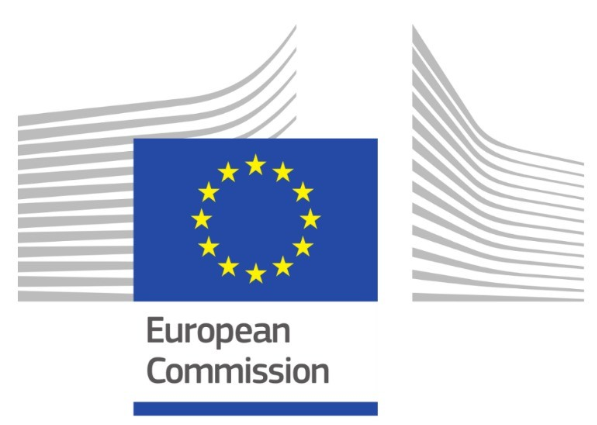}
\end{minipage}\hfil
\begin{minipage}{0.68\linewidth}

\noindent This research has received funding from the European Defence Industrial Development Programme (EDIDP) under the grant agreement Number EDIDP-CSAMN-SSC-2019-022-ECYSAP (European Cyber Situational Awareness Platform).
\end{minipage}


\bibliography{mybibfile}

\begin{thebibliography}{10}
\expandafter\ifx\csname url\endcsname\relax
  \def\url#1{\texttt{#1}}\fi
\expandafter\ifx\csname urlprefix\endcsname\relax\def\urlprefix{URL }\fi
\expandafter\ifx\csname href\endcsname\relax
  \def\href#1#2{#2} \def\path#1{#1}\fi

\bibitem{refMe-Sal4}
S.~Llopis~Sanchez, R.~Mazzolin, I.~Kechaoglou, D.~Wiemer, W.~Mees, J.~Muylaert,
  Handbook of Space Security, Springer, Cham, 2019, Ch. Cybersecurity Space
  Operation Center: Countering Cyber Threats in the Space Domain.
\newblock \href {http://dx.doi.org/10.1007/978-3-030-22786-9_108-1}
  {\path{doi:10.1007/978-3-030-22786-9_108-1}}.

\bibitem{newTDoS}
J.~Maestre~Vidal, M.~Sotelo~Monge, {{Denial of sustainability on military
  tactical clouds}}, in: Proceedings of the 15th International Conference on
  Availability, Reliability and Security (ARES), Dublin, Ireland, 2020, pp.
  1--9.

\bibitem{newCSA1}
D.~Sandoval Rodriguez-Bermejo, R.~Daton~Medenou, G.~Ramis Pasqual~de Riquelme,
  J.~Maestre~Vidal, F.~Torelli, S.~Llopis~Sánchez, {{Evaluation methodology
  for mission-centric cyber situational awareness capabilities}}, in:
  Proceedings of the 15th International Conference on Availability, Reliability
  and Security (ARES), Dublin, Ireland, 2020, pp. 1--9.

\bibitem{ares1}
D.~Raymond, T.~Cross, G.~Conti, M.~Nowatkowski, Key terrain in cyberspace:
  {Seeking} the high ground, in: 2014 6th {International} {Conference} {On}
  {Cyber} {Conflict} ({CyCon} 2014), IEEE, Tallin, Estonia, 2014, pp. 287--300,
  iSSN: 2325-5366.
\newblock \href {http://dx.doi.org/10.1109/CYCON.2014.6916409}
  {\path{doi:10.1109/CYCON.2014.6916409}}.

\bibitem{ares2}
{Giorgio Bertoli}, {Stephen Raio},
  \href{https://www.csiac.org/journal-article/the-elusive-nature-of-key-cyber-terrain/}{The
  elusive nature of “key cyber terrain”}, CSIAC Journal 6.
\newline\urlprefix\url{https://www.csiac.org/journal-article/the-elusive-nature-of-key-cyber-terrain/}

\bibitem{refMe-Sal1}
S.~Llopis, J.~Hingant, I.~Perez, M.~Esteve, F.~Carvajal, W.~Mees, T.~Debatty, A
  comparative analysis of visualisation techniques to achieve cyber situational
  awareness in the military, in: Proceedings of the 2018 International
  Conference on Military Communications and Information Systems, Warsaw,
  Poland, 2018, pp. 1--7.

\bibitem{newAresKCT}
A.~Luis~Martinez, V.~Villagra~Gonzalez, {{A novel automatic discovery system of
  critical assets in cyberspace-oriented military missions}}, in: Proceedings
  of the 15th International Conference on Availability, Reliability and
  Security (ARES), Dublin, Ireland, 2020, pp. 1--8.

\bibitem{ares4}
I.~T.~L. Computer Security~Division,
  \href{https://csrc.nist.gov/projects/risk-management/risk-management-framework-(RMF)-Overview}{Risk
  management framework (rmf) overview - fisma implementation project \textbar
  csrc}, library Catalog: csrc.nist.gov (Nov. 2016).
\newline\urlprefix\url{https://csrc.nist.gov/projects/risk-management/risk-management-framework-(RMF)-Overview}

\bibitem{refMe-new6}
J.~Maestre~Vidal, L.~Sotelo~Monge, M.A.and~Villalba, {A Novel Pattern
  Recognition System for Detecting Android Malware by Analyzing Suspicious Boot
  Sequences}, Knowledge-Based Systems 150 (2018) 198--217.

\bibitem{ares5}
{Joint Task Force Transformation Initiative},
  \href{https://nvlpubs.nist.gov/nistpubs/Legacy/SP/nistspecialpublication800-30r1.pdf}{Guide
  for conducting risk assessments}, Tech. Rep. NIST SP 800-30r1, National
  Institute of Standards and Technology, Gaithersburg, MD, edition: 0 (2012).
\newline\urlprefix\url{https://nvlpubs.nist.gov/nistpubs/Legacy/SP/nistspecialpublication800-30r1.pdf}

\bibitem{refNIst-stage1}
A.~Annarelli, F.~Nonino, G.~Palombi, Understanding the management of cyber
  resilient systems, Computers \& Industrial Engineering 149 (2020) 106829.

\bibitem{refNIst-stage2}
I.~Lee, Internet of things (iot) cybersecurity: Literature review and iot cyber
  risk management, Future Internet 12~(9).

\bibitem{refNIst-stage3}
R.~Knight, J.~Nurse, A framework for effective corporate communication after
  cyber security incidents, Computers \& Security 99 (2020) 102036.

\bibitem{refNIst-stage4}
L.~Barona~Lopez, A.~Valdivieso~Caraguay, J.~Maestre~Vidal, M.~Sotelo~Monge,
  L.~Villalba, Towards incidence management in 5g based on situational
  awareness, Future Internet 9~(1).

\bibitem{ares6}
R.~Caralli, J.~Stevens, L.~Young, W.~Wilson,
  \href{http://resources.sei.cmu.edu/library/asset-view.cfm?AssetID=8419}{Introducing
  octave allegro: Improving the information security risk assessment process},
  Tech. Rep. CMU/SEI-2007-TR-012, Software Engineering Institute, Carnegie
  Mellon University, Pittsburgh, PA (2007).
\newline\urlprefix\url{http://resources.sei.cmu.edu/library/asset-view.cfm?AssetID=8419}

\bibitem{ares7}
I.~262, \href{https://www.iso.org/obp/ui#iso:std:iso:31000:ed-2:v1:en}{{ISO}
  31000:2018 {Risk} management — {Guidelines}} (2018).
\newline\urlprefix\url{https://www.iso.org/obp/ui#iso:std:iso:31000:ed-2:v1:en}

\bibitem{ares8}
{ISO/IEC},
  \href{https://www.iso.org/cms/render/live/en/sites/isoorg/contents/data/standard/07/52/75281.html}{{ISO}/{IEC}
  27005:2018} (2018).
\newline\urlprefix\url{https://www.iso.org/cms/render/live/en/sites/isoorg/contents/data/standard/07/52/75281.html}

\bibitem{ares9}
ENISA,
  \href{https://www.enisa.europa.eu/topics/threat-risk-management/risk-management/current-risk/risk-management-inventory/rm-ra-methods/m_mehari.html}{Mehari}
  (2016).
\newline\urlprefix\url{https://www.enisa.europa.eu/topics/threat-risk-management/risk-management/current-risk/risk-management-inventory/rm-ra-methods/m_mehari.html}

\bibitem{ares10}
PAe,
  \href{https://administracionelectronica.gob.es/pae_Home/pae_Documentacion/pae_Metodolog/pae_Magerit.html}{{PAe}
  - {MAGERIT} v.3 : {Metodología} de {Análisis} y {Gestión} de {Riesgos} de
  los {Sistemas} de {Información}} (2012).
\newline\urlprefix\url{https://administracionelectronica.gob.es/pae_Home/pae_Documentacion/pae_Metodolog/pae_Magerit.html}

\bibitem{refQualitativeRM}
G.~Kavallieratos, S.~Katsikas, V.~Gkioulos, Cybersecurity and safety
  co-engineering of cyberphysical systems — a comprehensive survey, Future
  Internet 12~(4).

\bibitem{refQuantitativeRM-1}
A.~Rea-Guaman, J.~Mejia, T.~San~Feliu, J.~Calvo-Manzano, Avarciber: a framework
  for assessing cybersecurity risks, Cluster Computing 23 (2020) 1827--1843.

\bibitem{refQuantitativeRM-2}
L.~Gordon, M.~Loeb, L.~Zhou, Integrating cost–benefit analysis into the nist
  cybersecurity framework via the gordon-loeb model, Journal of Cybersecurity
  23.

\bibitem{ares11}
M.~Endsley, Situation awareness global assessment technique ({SAGAT}), in:
  Proceedings of the {IEEE} 1988 {National} {Aerospace} and {Electronics}
  {Conference}, Vol.~3, IEEE, Dayton, OH, USA, 1988, pp. 789--795.
\newblock \href {http://dx.doi.org/10.1109/NAECON.1988.195097}
  {\path{doi:10.1109/NAECON.1988.195097}}.

\bibitem{ares14}
G.~Bedny, D.~Meister, \href{https://doi.org/10.1207/s15327566ijce0301_5}{Theory
  of {Activity} and {Situation} {Awareness}}, International Journal of
  Cognitive Ergonomics 3~(1) (1999) 63--72, publisher: Routledge \_eprint:
  https://doi.org/10.1207/s15327566ijce0301\_5.
\newblock \href {http://dx.doi.org/10.1207/s15327566ijce0301_5}
  {\path{doi:10.1207/s15327566ijce0301_5}}.
\newline\urlprefix\url{https://doi.org/10.1207/s15327566ijce0301_5}

\bibitem{ares15}
K.~Smith, P.~A. Hancock,
  \href{https://doi.org/10.1518/001872095779049444}{Situation {Awareness} {Is}
  {Adaptive}, {Externally} {Directed} {Consciousness}}, Human Factors 37~(1)
  (1995) 137--148, publisher: SAGE Publications Inc.
\newblock \href {http://dx.doi.org/10.1518/001872095779049444}
  {\path{doi:10.1518/001872095779049444}}.
\newline\urlprefix\url{https://doi.org/10.1518/001872095779049444}

\bibitem{ares16}
M.~P.~T. Jr, Shaping and adapting: {Unlocking} the power of {Colonel} {John}
  {Boyd}’s {OODA} {Loop} (Apr. 2015).

\bibitem{ares17}
S.~Jajodia, S.~Noel, P.~Kalapa, M.~Albanese, J.~Williams, Cauldron:
  {Mission}-centric cyber situational awareness with defense in depth,
  Proceedings - IEEE Military Communications Conference MILCOM (2011)
  1339--1344\href {http://dx.doi.org/10.1109/MILCOM.2011.6127490}
  {\path{doi:10.1109/MILCOM.2011.6127490}}.

\bibitem{refMe-new7}
J.~Maestre~Vidal, A.~Orozco, L.~Villalba, {Adaptive artificial immune networks
  for mitigating DoS flooding attacks}, Swarm and Evolutionary Computation 38
  (2018) 94--108.

\bibitem{refMe-Sal3}
K.~Demertzis, N.~Tziritas, P.~Kikiras, S.~Llopis~Sanchez, L.~Iliadis, {The Next
  Generation Cognitive Security Operations Center: Network Flow Forensics Using
  Cybersecurity Intelligence}, Big Data and Cognitive Computing 2~(35).

\bibitem{refMe-new3}
J.~Maestre~Vidal, M.~Sotelo~Monge, { Obfuscation of Malicious Behaviors for
  Thwarting Masquerade Detection Systems Based on Locality Features}, Sensors
  20(7)~(2084).

\bibitem{refMe-new1}
M.~A. Sotelo~Monge, J.~Maestre~Vidal, G.~Mart{\'\i}nez~P{\'e}rez, Detection of
  economic denial of sustainability (edos) threats in self-organizing networks,
  Computer Communications 145 (2019) 284--308.

\bibitem{refMe-Sal2}
K.~Demertzis, N.~Tziritas, P.~Kikiras, S.~Llopis~Sanchez, L.~Iliadis, {The Next
  Generation Cognitive Security Operations Center: Adaptive Analytic Lambda
  Architecture for Efficient Defense against Adversarial Attacks}, Big Data and
  Cognitive Computing 3~(6).

\bibitem{refMe-new5}
J.~Maestre~Vidal, M.~Sotelo~Monge, {Framework for Anticipatory Self-Protective
  5G Environments}, in: Proceedings of the 14th nternational Conference on
  Availability, Reliability and Security (ARES).

\bibitem{refMe-new8}
J.~Maestre~Vidal, M.~Sotelo~Monge, {A novel Self-Organizing Network solution
  towards Crypto-ransomware Mitigation}, in: Proceedings of the 13th
  nternational Conference on Availability, Reliability and Security (ARES).

\bibitem{ares18}
{McGuinness, B.}, {Foy, L.}, A {Subjective} {Measure} of {SA}: {The} {Crew}
  {Awareness} {Rating} {Scale} ({CARS}), in: 1st, {Human} performance,
  situation awareness and automation conference; user-centered design for the
  new millennium, SA Technologies, Savannah, GA, 2000, p.~6.

\bibitem{ares19}
V.~Lenders, A.~Tanner, A.~Blarer, Gaining an {Edge} in {Cyberspace} with
  {Advanced} {Situational} {Awareness}, IEEE Security Privacy 13~(2) (2015)
  65--74, conference Name: IEEE Security Privacy.
\newblock \href {http://dx.doi.org/10.1109/MSP.2015.30}
  {\path{doi:10.1109/MSP.2015.30}}.

\bibitem{ares20}
D.~Buckshaw, G.~Parnell, W.~Unkenholz, D.~Parks, J.~Wallner, O.~Saydjari,
  Mission {Oriented} {Risk} and {Design} {Analysis} of {Critical} {Information}
  {Systems}, Military Operations Research 10 (2005) 19--38.
\newblock \href {http://dx.doi.org/10.5711/morj.10.2.19}
  {\path{doi:10.5711/morj.10.2.19}}.

\bibitem{ares21}
D.~S.~E. Noel, W.~J. Heinbockel,
  \href{https://www.mitre.org/publications/technical-papers/an-overview-of-mitre-cyber-situational-awareness-solutions}{An
  {Overview} of {MITRE} {Cyber} {Situational} {Awareness} {Solutions}}, in:
  NATO Cyber Defence Situational Awareness Solutions Conference, MITRE,
  Bucharest, Romania, 2015, p.~17.
\newline\urlprefix\url{https://www.mitre.org/publications/technical-papers/an-overview-of-mitre-cyber-situational-awareness-solutions}

\bibitem{ares22}
J.~Hingant~Gómez, M.~Zambrano, F.~Pérez, I.~Perez-Llopis, M.~Esteve,
  {HYBINT}: {A} {Hybrid} {Intelligence} {System} for {Critical}
  {Infrastructures} {Protection}, Security and Communication Networks 2018
  (2018) 1--13.
\newblock \href {http://dx.doi.org/10.1155/2018/5625860}
  {\path{doi:10.1155/2018/5625860}}.

\bibitem{ares23}
U.~Franke, J.~Brynielsson,
  \href{http://www.sciencedirect.com/science/article/pii/S0167404814001011}{Cyber
  situational awareness – {A} systematic review of the literature}, Computers
  \& Security 46 (2014) 18--31.
\newblock \href {http://dx.doi.org/10.1016/j.cose.2014.06.008}
  {\path{doi:10.1016/j.cose.2014.06.008}}.
\newline\urlprefix\url{http://www.sciencedirect.com/science/article/pii/S0167404814001011}

\bibitem{ares24}
{Raymond T. Odierno}, Field {Manual} 3-90-1: {Offense} and {Defense}, Tech.
  Rep. FM 3-90-1, Department of the Army (Mar. 2013).

\bibitem{ares25}
N.~T. Pantin, \href{https://calhoun.nps.edu/handle/10945/53030}{Key terrain:
  application to the layers of cyberspace}, Thesis, Naval Postgraduate School,
  Monterey, California, accepted: 2017-05-10T16:31:58Z (Mar. 2017).
\newline\urlprefix\url{https://calhoun.nps.edu/handle/10945/53030}

\bibitem{ares27}
{National Park Service}, {OCOKA} {Military} {Terrain} {Analysis}.

\bibitem{ares28}
D.~Raymond, G.~Conti, T.~Cross, R.~Fanelli, A control measure framework to
  limit collateral damage and propagation of cyber weapons, in: 2013 5th
  {International} {Conference} on {Cyber} {Conflict} ({CYCON} 2013), IEEE,
  Tallinn, Estonia, 2013, pp. 1--16, iSSN: 2325-5366.

\bibitem{ares29}
P.~Price, N.~Leyba, M.~Gondree, Z.~Staples, T.~Parker, Asset {Criticality} in
  {Mission} {Reconfigurable} {Cyber} {Systems} and its {Contribution} to {Key}
  {Cyber} {Terrain}, in: Proceedings of the 50th {Hawaii} {International}
  {Conference} on {System} {Sciences}, Honolulu, Hawaii, 2017, p.~10.
\newblock \href {http://dx.doi.org/10.24251/HICSS.2017.729}
  {\path{doi:10.24251/HICSS.2017.729}}.

\bibitem{ares30}
B.~Thompson, R.~Harang,
  \href{https://doi.org/10.1145/3041008.3041015}{Identifying {Key}
  {Cyber}-{Physical} {Terrain}}, in: Proceedings of the 3rd {ACM} on
  {International} {Workshop} on {Security} {And} {Privacy} {Analytics}, {IWSPA}
  '17, Association for Computing Machinery, Scottsdale, Arizona, USA, 2017, pp.
  23--28.
\newblock \href {http://dx.doi.org/10.1145/3041008.3041015}
  {\path{doi:10.1145/3041008.3041015}}.
\newline\urlprefix\url{https://doi.org/10.1145/3041008.3041015}

\bibitem{ares31}
A.~E. Schulz, M.~C. Kotson, J.~R. Zipkin,
  \href{https://apps.dtic.mil/docs/citations/ADA635942}{Cyber {Network}
  {Mission} {Dependencies}}, Tech. Rep. TR-1189, Massachusetts Institute of
  Technology Lexington Lincoln Lab, Lexington, Massachusetts (Sep. 2015).
\newline\urlprefix\url{https://apps.dtic.mil/docs/citations/ADA635942}

\bibitem{refMissionPlanner}
X.~{Li}, N.~{Bo}, W.~{Pang}, J.~{Tang}, Software architecture of c2 mission
  system based on mas and soa for manned/unmanned aerial vehicle formation, in:
  Proc. of the 8th IEEE International Conference on Software Engineering and
  Service Science (ICSESS), 2017, pp. 842--845.

\bibitem{refBPMN}
L.~Sabatucci, M.~Cossentino, Supporting dynamic workflows with automatic
  extraction of goals from bpmn, ACM Transactions on Autonomous and Adaptive
  Systems 14~(2).

\bibitem{refSISO}
S.~I.~S. Organization,
  \href{https://www.sisostds.org/ProductsPublications/Standards/SISOStandards.aspx}{Standard
  for military scenario definition language (siso-std-007-2008)} (2008).
\newline\urlprefix\url{https://www.sisostds.org/ProductsPublications/Standards/SISOStandards.aspx}

\bibitem{ref5gmodel}
M.~Sotelo~Monge, J.~Maestre~Vidal, L.~Villalba, Reasoning and knowledge
  acquisition framework for 5g network analytics, Sensors 17~(10).

\bibitem{refMonitoring}
A.~{D’Alconzo}, I.~{Drago}, A.~{Morichetta}, M.~{Mellia}, P.~{Casas}, A
  survey on big data for network traffic monitoring and analysis, IEEE
  Transactions on Network and Service Management 16~(3) (2019) 800--813.

\bibitem{refNoC}
C.~{Onwubiko}, Cyber security operations centre: Security monitoring for
  protecting business and supporting cyber defense strategy, in: Proc. of the
  2015 International Conference on Cyber Situational Awareness, Data Analytics
  and Assessment (CyberSA), 2015, pp. 1--10.

\bibitem{refNew-CTIGathering}
W.~Tounsi, H.~Rais, A survey on technical threat intelligence in the age of
  sophisticated cyber attacks, Computers \& Security 72 (2018) 212--233.

\bibitem{new-hybridDamage}
{{Multinational Capability Development Campaign project}}, Countering hybrid
  warfare project:understanding hybrid warfare,
  \url{https://www.gov.uk/government/publications/countering-hybrid-warfare-project-understanding-hybrid-warfare}
  (2017).

\bibitem{ref-PlataformaDato}
A.~{Munshi}, Y.~{Mohamed}, Data lake lambda architecture for smart grids big
  data analytics, IEEE Access 6 (2018) 40463--40471.

\bibitem{ref-newDCS}
J.~van~der Geest, C.~van~den Broek, H.~Bastiaansen, M.~Schenk, {{Enabling a Big
  Data and AI Infrastructure with a Data Centric and Microservice Approach:
  Challenges and Developments}}, in: Procceedings of the NATO Specialist
  Meeting on Big Data \& Artificial Intelligence for Military Decision Making,
  Bordeaux, France, 2019, pp. 1--9.

\bibitem{ref-newOODA}
B.~Brehmer, {{The dynamic OODA loop: Amalgamating Boyd's OODA loop and the
  cybernetic approach to command and control}}, in: Proceedings of the 10th
  International Command and Control Research and Technology Symposium, MacLean,
  Virginia, USA, 2005, pp. 365--368.

\bibitem{ares32}
MITRE, \href{https://cpe.mitre.org/specification/}{{CPE} - {Common} {Platform}
  {Enumeration}: {CPE} {Specifications}} (2013).
\newline\urlprefix\url{https://cpe.mitre.org/specification/}

\bibitem{ref-KOCOA}
D.~Spennemann, Using kocoa military terrain analysis for the assessment of
  twentieth century battlefield landscapes, Heritage 3~(3) (2020) 753--781.

\bibitem{newTDoS94}
{NATO}, Allied joint doctrine for cyberspace operation (ajp-3.20).

\bibitem{ares33}
Y.~Nir, R.~Salz, N.~Sullivan,
  \href{https://www.iana.org/assignments/tls-parameters/tls-parameters.xhtml}{Transport
  {Layer} {Security} ({TLS}) {Parameters}} (2020).
\newline\urlprefix\url{https://www.iana.org/assignments/tls-parameters/tls-parameters.xhtml}

\bibitem{ref-hilltop}
S.~Applegate, C.~Carpenter, D.~West, Searching for digital hilltops, Joint
  Force Quarterly 84~(1) (2017) 18--23.

\bibitem{dompfi}
T.~{Tuukkanen}, S.~{Couturier}, B.~{Buchin}, T.~{Bräysy}, J.~{Krygier},
  E.~{Verheul}, V.~{Le Nir}, N.~{Smit}, Assessment of cognitive radio networks
  through military capability development viewpoint, in: Proceedings og the
  International Conference on Military Communications and Information Systems
  (ICMCIS), 2018, pp. 1--8.

\bibitem{ref-ThreeSwords}
A.~Rizwan, on cyber defence, The Three Swords Magazine 24 (2014) 32--47.

\bibitem{newDoctrine-offensive}
{US Army Doctrine Publication}, Offensive and defensive (3-90).

\bibitem{ref-capabilities-organization}
M.~G. {Jaatun}, L.~{Bodsberg}, T.~O. {Grøtan}, M.~{Elisabeth Gaup Moe}, An
  empirical study of cert capacity in the north sea, in: Proceedings of the
  2020 International Conference on Cyber Security and Protection of Digital
  Services (Cyber Security), 2020, pp. 1--8.

\bibitem{ref-capabilities-nice}
W.~Newhouse, S.~Keith, B.~Scribner, G.~Witte, National initiative for
  cybersecurity education (nice) cybersecurity workforce framework, NIST
  special publication~(800).

\bibitem{FURNELL20206}
S.~Furnell, M.~Bishop, Addressing cyber security skills: the spectrum, not the
  silo, Computer Fraud \& Security 2020~(2) (2020) 6--11.

\bibitem{ZHOU20189}
D.~Zhou, Z.~Yan, Y.~Fu, Z.~Yao, A survey on network data collection, Journal of
  Network and Computer Applications 116 (2018) 9--23.

\bibitem{ZARPELAO201725}
B.~Zarpelao, R.~Miani, C.~Kawakani, S.~Carlisto, A survey of intrusion
  detection in internet of things, Journal of Network and Computer Applications
  84 (2017) 25--37.

\bibitem{BUCHLER2018114}
N.~Buchler, P.~Rajivan, L.~Marusich, L.~Lightner, C.~Gonzalez, Sociometrics and
  observational assessment of teaming and leadership in a cyber security
  defense competition, Computers \& Security 73 (2018) 114--136.

\bibitem{ref-capabilities-doctrine}
M.~Taddeo, L.~Floridi, Regulate artificial intelligence to avert cyber arms
  race, Nature 556 (2019) 296--298.

\bibitem{doi:10.1080/14702436.2015.1108108}
J.~Burton, Nato’s cyber defence: strategic challenges and institutional
  adaptation, Defence Studies 15~(4) (2015) 297--319.

\bibitem{ref-capabilities-facilities}
J.~{Kukkola}, Civilian and military information infrastructure and the control
  of the russian segment of internet, in: Proceedings of the 2018 International
  Conference on Military Communications and Information Systems (ICMCIS), 2018,
  pp. 1--8.

\end{thebibliography}

\end{document}